\documentclass[
    aps,
    prresearch,              % or pra, prl, rmp, etc.
    reprint,          % APS two-column journal style
    superscriptaddress,
    longbibliography,
    nofootinbib
]{revtex4-2}

\pdfoutput=1

\usepackage{amsthm,amsfonts,mathrsfs,mathdots,graphicx,times,xfrac,mathtools,enumitem,xr,bbm,verbatim,dsfont,soul,array,multirow, comment, physics}

\usepackage{placeins}

\usepackage{tikz}
\usepackage{lipsum}

\usepackage[utf8]{inputenc}
\usepackage[english]{babel}
\usepackage[T1]{fontenc}
\usepackage{amsmath}
\usepackage{hyperref}

\usepackage{listings}
\usepackage{xcolor}
\usepackage{physics}
\usepackage{amsmath}
\usepackage{amssymb} 
\usepackage{float}
\usepackage[caption=false]{subfig}

\newcommand\scalemath[2]{\scalebox{#1}{\mbox{\ensuremath{\displaystyle #2}}}}
\newcommand{\orcid}[1]{}

\begin{document}

\title{Closed-Form Generators for Perturbative Transformations in Static and Periodically Driven Quantum Systems
}

\author{Leander Reascos}
\email{irving.reascos.valencia@uni-a.de}
\affiliation{Augsburg University, Institute of Physics, Universitätsstraße 1 (Physik Nord), 86159 Augsburg}
\orcid{0009-0008-0141-2783}

\author{Giovanni Francesco Diotallevi}
\email{francesco.diotallevi@uni-a.de}
\affiliation{Augsburg University, Institute of Physics, Universitätsstraße 1 (Physik Nord), 86159 Augsburg}
\orcid{0000-0001-6765-7806}

\author{Mónica Benito}
\affiliation{Augsburg University, Institute of Physics, Universitätsstraße 1 (Physik Nord), 86159 Augsburg}

    \affiliation{Center for Advanced Analytics and Predictive Sciences, University of Augsburg, Augsburg, Germany}
\orcid{0000-0001-6365-9505}

\makeatletter
\newcommand{\equalcontrib}{%
  \begingroup
  \renewcommand\thefootnote{}\footnotetext{$^{\star,\dagger}$ These authors contributed equally to this work.}
  \endgroup
}
\makeatother

\begin{abstract}
\noindent
We introduce a unified framework for the practical implementation of perturbative 
transformations that overcomes the limitations of existing construction methods.
%limitations surrounding their construction. 
The central result is a closed-form operator expression for the generator of a broad class of perturbative transformations in systems composed of finite-dimensional and bosonic subspaces. We further generalize the construction to time-dependent systems with periodic perturbations, obtaining a solution that remains valid across low-, intermediate-, and high-frequency regimes, provided that the transition channels eliminated by the transformation remain sufficiently detuned. The derived framework remains independent of the particular perturbative scheme and therefore applies to a large class of perturbative approaches for both time independent and time dependent transformations. We demonstrate the scope and accuracy of the method by deriving effective dispersive interactions in exemplary anharmonic and periodically driven light--matter systems.
\end{abstract}

\maketitle

\equalcontrib

%%%%%%%%%%%%%%%%%%%%%%%%%%%%%%%%%%%%%%%%%%%%%%%%
%%%%%%%%%%%%%%%%%%%%%%%%%%%%%%%%%%%%%%%%%%%%%%%%
\section{Introduction}\label{sec: introduction}

Perturbative unitary transformations constitute a broad class of operator-based methods for deriving effective descriptions of quantum systems. Rather than determining perturbative corrections to individual eigenstates or energy levels, these transformations reorganize the Hamiltonian through a near-identity unitary rotation, allowing selected interactions to be eliminated or transferred into effective terms order by order. This operator-level perspective is particularly valuable when the objective is not only to approximate the spectrum, but also to identify the physical processes generated by virtual transitions, such as effective couplings, dispersive shifts, exchange interactions, and state-dependent nonlinearities.

The Schrieffer--Wolff transformation (SWT), also known as Löwdin perturbation theory~\cite{Lodwin_PT}, is the canonical example of this broader family of methods. It is conventionally employed to decouple energetically separated subspaces and construct an effective Hamiltonian acting within the sector of interest. This approach has become an established tool across condensed matter physics~\cite{SW_fermi_hubbard_model,SW_ising_quantum_criticality,SW_kondo_lattice,SW_mott_insulators,SW_quantum_lattice_models,SW_superconductor_physics_1,SW_superconductive_physics_2,SW_XXZ_chain}, quantum optics~\cite{SW_chain_coupled_to_cavity,SW_dissipative_bosonic_modes,SW_microwave_theory,SW_optical_lattice_interacting_bosons,SW_spin_squeezed_resonator}, and quantum information science~\cite{SW_electrons_in_helium,SW_anomalous_zero_field_splitting,SW_electrical_control_of_hole_qubits,SW_cavity_control_of_spin_qubits,SW_dispersive_two_qubit_gates,SW_emergent_linear_rashba_in_holes,SW_entanglement_generation,SW_hole_spin_qubits_in_Si_FinFETs,SW_hole_spin_qubits_in_wells,SW_Light_hole_spin_qubit,SW_quantum_algorithms,SW_quantum_algorithms_1,SW_quantum_phase_estimation_algorithm,SW_recent_advances_in_hole_spin_qubits}, as well as in dissipative, non-Hermitian, and driven settings~\cite{SW_dissipative_systems,SW_dynamic_nuclear_polarization,SW_krein_hermitian_hamiltonians,SW_non_hermitian_systems,SW_scar_states}. More generally, the usefulness of the SWT lies in its ability to expose the relevant physics while retaining the operator structure of the original problem.

This perspective extends beyond the conventional separation into two energy sectors. Modified SWT schemes, multi-block diagonalization procedures, and recursive transformations demonstrate that perturbative unitary methods can be adapted to eliminate different classes of couplings or reorganize a Hamiltonian according to objectives other than the construction of a low-energy theory~\cite{SW_MultiblockDiagonalization,pymablock,DiVincenzo_multiblock,SW_RecursiveSW}. Although these schemes differ in the terms they aim to eliminate, they share the same central technical problem: the construction of the anti-Hermitian generator that implements the desired transformation at each perturbative order.

Existing approaches to the construction of the generator can be classified into operator-level and matrix-element-based methods. In operator-level treatments, the generator is determined via a guess ansatz aimed at solving a commutator equation involving the unperturbed Hamiltonian and the terms targeted for elimination. This preserves the algebraic structure of the problem and often provides a transparent interpretation of the effective interactions. In practice, however, the required operator ansatz is frequently inferred from the particular form of the Hamiltonian~\cite{SW_superconductor_physics_Ansatz,SW_transferring_entangled_states_Ansatz,SW_beyond_dispersive_regime_Ansatz,SW_effective_hopping_in_silicene_Ansatz,SW_photodriven_germanium_hole_qubit_Ansatz,SW_role_of_anisotropic_confining_of_Ge_qubit_Ansatz}. Such constructions can become cumbersome for systems containing several distinct transition processes, explicit time dependencies, or infinite-dimensional degrees of freedom, and their generalization to new models may require substantial system-specific reasoning.

Matrix-element-based approaches offer a complementary route by expressing the generator directly in terms of transitions between eigenstates of the unperturbed Hamiltonian~\cite{SW_wikler_book,SW_quantum_many_body_systems_spin_lattices,SW_quantum_computation_protocol_matrixbased,SW_super_conducting_qubits_matrixbased,SW_ultrafast_Ge_qubits_matrixbased,SW_flopping_mode_qubits}. These methods are systematic and avoid the need to guess an operator ansatz, but they tend to obscure the operator structure responsible for the resulting effective interactions. Their application to systems with infinite-dimensional Hilbert spaces also generally requires a numerical truncation, thereby replacing an exact operator construction with a finite-dimensional approximation whose convergence must be assessed independently. The operator and matrix-element viewpoints therefore possess complementary advantages, yet neither by itself provides a fully general and readily applicable solution for composite systems containing both finite- and infinite-dimensional subspaces.

Several works have sought to improve the generality of the generator construction by introducing integral representations, eigenoperator decompositions, or closed-form expressions for particular classes of Hamiltonians~\cite{SW_quantum_information_processing,2021_closed_form,landi2024eigenoperatorapproachschriefferwolffperturbation}. These developments considerably reduce the amount of model-specific algebra required in suitable settings. Nevertheless, existing closed-form treatments are commonly restricted by the structure of the underlying Hilbert space, the form of the unperturbed Hamiltonian, or the assumption of time-independent perturbations. In particular, a general operator construction that simultaneously accommodates a finite-dimensional subsystem, an untruncated bosonic sector, possible anharmonicities, and explicitly time-dependent perturbations remains (to our knowledge) lacking.

In this work, we develop such a framework for a broad class of perturbative unitary transformations. Our central result is a closed-form operator construction of the generator that accommodates all the aforementioned limitations. 
 As a result, it eliminates the need for a system-specific ansatz, truncation, or additional calculations of the generator.
%As a result, no system-specific ansatz, truncation, nor additional calculations for the generator are required.
The construction is also independent of the particular terms chosen for elimination, and is therefore applicable not only to conventional perturbative methods (e.g. the SWT), but more generally to any perturbative unitary scheme.
%%%%%%%%%%%%%%%%%%%%%%%%%%%%%%%%%%%%%%%%%%%%%%%%
%%%%%%%%%%%%%%%%%%%%%%%%%%%%%%%%%%%%%%%%%%%%%%%%
 
\section{Framework} \label{sec: framework}
%%%%%%%%%%%%%%%%%%%%%%%%%%%%%%%%%%%%%%%%%%%%%%%%
\subsection{Static Perturbative Transformations} \label{sec: perturbative unitary transformations}

Consider a general quantum system described by a Hamiltonian $\hat{H}\equiv \hat{H}(\lambda=1)$ that admits a perturbative expansion in a dimensionless bookkeeping parameter $\lambda$,
\begin{align}
    \hat{H}(\lambda)
    =
    \hat{H}^{(0)}
    +
    \sum_{j=1}^{\infty}
    \lambda^j \hat{H}^{(j)} ,
    \label{eq: separated hamiltonian}
\end{align}
where $\hat{H}^{(0)}$ represents a solvable reference Hamiltonian whose eigenstates $\{\ket{n}\}$ and eigenvalues $\{E_n\}$ are known. The operators $\hat{H}^{(j)}$ denote contributions of perturbative order $j$ in $\lambda$, which may either couple different sectors of the spectrum or provide perturbative corrections to the bare energies. The physical Hamiltonian is recovered by setting $\lambda=1$ after the perturbative order counting has been performed.

The central idea underlying perturbative unitary transformations is to construct a unitary operator $\hat{U} = e^{-\hat{S}}$ that rotates the Hamiltonian into a simpler form. The transformed Hamiltonian (from here on we denote unitarily transformed effective Hamiltonians by calligraphic $\hat{\mathcal H}$)  
\begin{align}
\hat{\mathcal{H}} = \hat{U} \hat{H} \hat{U}^\dagger = e^{-\hat{S}} \hat{H} e^{\hat{S}} \label{eq: complete time independent rotation}
\end{align}
encodes the same physics as the original system but in a basis where certain coupling terms are systematically reduced or eliminated. The generator $\hat{S}$ is chosen to be anti-Hermitian, $\hat{S}^\dagger = -\hat{S}$, ensuring that $\hat{U}$ remains unitary; thus, this preserves the spectrum of the full Hamiltonian before any perturbative truncation is made.

Expanding the transformation via the Baker-Campbell-Hausdorff formula yields 
\begin{align}
\hat{\mathcal{H}} = \hat{H} + [\hat{H}, \hat{S}] + \frac{1}{2}[[\hat{H}, \hat{S}], \hat{S}] + \cdots,
\end{align}
a series whose terms organize naturally according to the perturbative order of $\hat{S}$. If we write $\hat{S} = \sum_{j=1} \lambda^j \hat{S}^{(j)}$ as a perturbative expansion matching the structure of our Hamiltonian, we can systematically construct $\hat{\mathcal{H}}$ order by order. At second order, for instance, this expansion becomes

\begin{align}
    \nonumber \hat{\mathcal{H}} \approx& \sum_{i=0}^2 \lambda^i\hat{H}^{(i)} + \lambda[\hat{H}^{(0)}, \hat{S}^{(1)}] +  \lambda^2\left([\hat{H}^{(0)}, \hat{S}^{(2)}]\right.+\\
     & \left.+  [\hat{H}^{(1)}, \hat{S}^{(1)}] + \frac{1}{2}[[\hat{H}^{(0)}, \hat{S}^{(1)}], \hat{S}^{(1)}]\right). \label{eq: expanded H_tilde}
\end{align}

\begin{figure*}[t]
    \centering
    \includegraphics[width=0.92\textwidth]{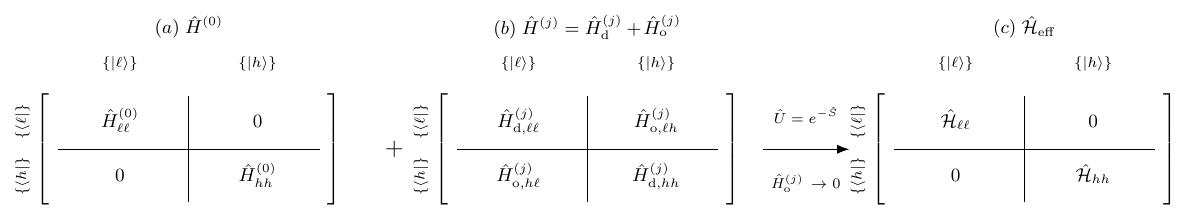}
    \caption{
    Schematic block structure of the perturbative decomposition used in the separation of energetically distant subspaces.
    The unperturbed Hamiltonian $\hat H^{(0)}$ is block diagonal with respect to the low- and high-energy subspaces $\{\ket{\ell}\}$ and $\{\ket{h}\}$. At perturbative order $j$, the correction $\hat H^{(j)}$ is decomposed into a block-diagonal contribution $\hat H_{\rm d}^{(j)}$, acting within each subspace, and an off-diagonal contribution $\hat H_{\rm o}^{(j)}$, coupling the two subspaces. The unitary transformation $\hat U=e^{-\hat S}$ is chosen such that the off-diagonal coupling is eliminated
    to the desired perturbative order, yielding an effective block-diagonal Hamiltonian
    $\hat{\mathcal H}_{\rm eff}$.
    }
    \label{fig:SWT_block_separation}
\end{figure*}

The power of this framework lies in its flexibility. By choosing different targets for $\hat{\mathcal{H}}$, we obtain different perturbative schemes, each suited to particular physical contexts. The defining equation for the generator $\hat{S}^{(j)}$ takes the general form
\begin{align}
    \left[\hat{H}^{(0)}, \hat{S}^{(j)}\right] = \hat{P}^{(j)}, \label{eq: Equation to solve}
\end{align}
where, going back to the second order expansion example, the operator $\hat{P}^{(j)}$ represents whichever combination of terms from Eq.~\eqref{eq: expanded H_tilde} we wish to eliminate at order $j$. Different perturbative schemes differ primarily in the choice of the target operator $\hat P^{(j)}$; however, the algebraic task can always be reframed as the inversion of the commutator presented in Eq.~\eqref{eq: Equation to solve}.

The SWT emerges as a particularly important special case of this framework. When the spectrum of $\hat{H}^{(0)}$ naturally divides into two well-separated subspaces, with low-energy states labelled as $\{\ket{\ell}\}$ and high-energy states $\{\ket{h}\}$, we can choose $\hat{P}^{(j)}$ to eliminate the off-diagonal blocks that couple these subspaces. To be more specific, consider a Hamiltonian system in which perturbative corrections can be further decomposed as 
\begin{align}
    \hat{H}^{(j)} = \hat{H}^{(j)}_{\rm d} + \hat{H}^{(j)}_{\rm o},
\end{align}
where $\hat{H}^{(j)}_{\rm d}$ represents perturbative interactions and (or) diagonal corrections acting solely within the block diagonal subspaces defined by the $\{\ket{\ell}\}$ or $\{\ket{h}\}$ states, while $\hat{H}^{(j)}_{\rm o}$ represent perturbative interactions between the two block subspaces, as illustrated in Fig.~\ref{fig:SWT_block_separation}.
Setting
\begin{align}
    \hat{P}^{(1)} &= -\hat{H}_{\rm o}^{(1)}, \label{eq: equation to solve s1}\\
    \hat{P}^{(2)} &= -\hat{H}_{\rm o}^{(2)} - [\hat{H}_{\rm d}^{(1)}, \hat{S}^{(1)}], \label{eq: equation to solve s2}
\end{align}
ensures that $\hat{\mathcal{H}}$ becomes block-diagonal with respect to the $\{\ket{\ell}\}$ and $\{\ket{h}\}$ subspaces to second order in perturbation theory. This block diagonalization allows us to analyze the low-energy physics while systematically incorporating the influence of high-energy states through effective interactions, making it invaluable for studying phenomena in condensed matter physics, quantum optics, and quantum information science where disparate energy scales naturally emerge.

The historical development of methods for solving Eq.~\eqref{eq: Equation to solve} has followed two complementary paths. A matrix-element approach, which offers a systematic procedure at the cost of working in a fixed basis. By computing the matrix elements of Eq.~\eqref{eq: Equation to solve} between basis states, one arrives at an explicit formula~\cite{SW_wikler_book}
\begin{align}
\bra{\ell}\hat{S}^{(j)}\ket{h} = \frac{\bra{\ell}\hat{P}^{(j)}\ket{h}}{E_\ell - E_h},\quad
E_\ell\neq E_h.
\label{eq: sum over indices approach}
\end{align}
This expression directly connects the generator to the perturbation through the inverse energy differences, capturing the intuition that strong perturbations between nearly degenerate states require careful treatment. However, this approach sacrifices operator-level insights and becomes problematic for infinite-dimensional systems, which require truncation of the Hilbert space; a procedure that can compromise accuracy if not performed with care.

The alternative  operator-level approach preserves the algebraic structure of the problem. Here, one attempts to construct $\hat{S}^{(j)}$ directly by identifying operator patterns that satisfy Eq.~\eqref{eq: Equation to solve}. While this method maintains physical transparency and avoids the need to truncate infinite-dimensional spaces, it often relies on inspired guesswork that becomes increasingly difficult for complex perturbations. Along these lines, a solution presented in Ref.~\cite{SW_quantum_information_processing}, provides an elegant integral formulation
\begin{align} 
\hat{S}^{(j)} = \lim_{t^\prime \to 0} \left[i\int e^{i\hat{H}^{(0)}t^\prime} \hat{P}^{(j)}e^{-i\hat{H}^{(0)}t^\prime}  dt^\prime\right]. \label{eq: interaction picture solution}
\end{align}
This expression interprets the generator as the accumulated effect of the perturbation in the interaction picture, offering both mathematical rigor and physical insight. Yet its practical implementation still requires system-specific calculations, and it stops short of providing a fully closed-form solution.

%%%%%%%%%%%%%%%%%%%%%%%%%%%%%%%%%%%%%%%%%%%%%%%%
\subsection{Time-Dependent Perturbative Transformations}\label{sec: time dependent transformations}

The framework developed above naturally extends to systems where the Hamiltonian carries explicit time dependence, a situation ubiquitous in modern quantum technologies. External control fields, parametric driving, and modulated couplings introduce time-varying perturbations that can dramatically alter system behavior, enabling phenomena from dynamical decoupling to parametric amplification, and from Floquet engineering of synthetic gauge fields to the coherent manipulation of quantum states. The question becomes: how do we generalize our perturbative transformation scheme to capture these time-dependent effects while maintaining the systematic structure presented above?

When the Hamiltonian becomes time-dependent, $\hat{H} \to \hat{H}(t)$, our transformation must adapt accordingly. The generator $\hat{S}$ now carries time dependence, $\hat{S} \to \hat{S}(t)$, and the effective Hamiltonian emerges from the time-dependent unitary rotation~\cite{SW_TD_subharmonic_transition_and_block_siegert, Brooks_Burkard_TDSW_2020}
\begin{align}
\hat{\mathcal{H}}(t) = e^{-\hat{S}(t)} \hat{H}(t) e^{\hat{S}(t)} + i\hbar \frac{\partial e^{-\hat{S}(t)}}{\partial t} e^{\hat{S}(t)}. \label{eq: complete time dependent rotation}
\end{align}
The crucial difference from the static case in Eq.~\eqref{eq: complete time independent rotation} is the appearance of the second term on the right-hand side. This contribution arises because the transformation itself evolves in time, introducing an additional energy associated with the rate of change of the basis. Physically, this term represents the non-adiabatic corrections that emerge when the transformation cannot instantaneously follow the time-dependent perturbation.

The time derivative term in Eq.~\eqref{eq: complete time dependent rotation} can be expanded as~\cite{SW_TD_subharmonic_transition_and_block_siegert, Brooks_Burkard_TDSW_2020}
\begin{align}
    \frac{\partial e^{-\hat{S}(t)}}{\partial t} e^{\hat{S}(t)} = \sum_{l=0}^\infty \frac{-1}{(l+1)!} \left[\frac{\partial \hat{S}(t)}{\partial t}, \hat{S}(t)\right]^{(l)},
\end{align}
where the notation $[\cdot, \cdot]^{(l)}$ denotes nested commutators of order $l$. This expansion reveals that the time derivative of $\hat{S}(t)$ generates an infinite series of corrections through its commutation relations with $\hat{S}(t)$ itself. Whether these corrections matter depends critically on the magnitude of $\frac{\partial \hat{S}^{(j)}(t)}{\partial t}$ relative to the perturbative order of $\hat{S}^{(j)}(t)$.

Two distinct regimes emerge from this consideration. In the first regime, the transformation varies slowly enough that $\frac{\partial \hat{S}^{(j)}(t)}{\partial t}$ is of higher perturbative order than $\hat{S}^{(j)}(t)$ itself. This occurs when the characteristic timescale of variation in the perturbation is much longer than the relevant dynamical timescales of the unperturbed system. In such cases, the transformation can adiabatically follow the time-dependent perturbation, and the derivative term contributes only at higher orders than our working accuracy. The static formalism then carries through essentially unchanged, with the understanding that all operators now carry parametric time dependence: $\hat{S}^{(j)} \to \hat{S}^{(j)}(t)$ and $\hat{P}^{(j)} \to \hat{P}^{(j)}(t)$, while the defining equation Eq.~\eqref{eq: Equation to solve} maintains its form.

The second regime, far more subtle and physically rich, occurs when the time variation of $\hat{S}^{(j)}(t)$ contributes at the same perturbative order as $\hat{S}^{(j)}(t)$ itself. This happens when external drives or parametric modulations oscillate at frequencies comparable to, or faster than, the internal energy scales of the system. In this non-adiabatic regime, the derivative term in Eq.~\eqref{eq: complete time dependent rotation} cannot be neglected, and the defining equation for the generator acquires an additional contribution. Specifically, Eq.~\eqref{eq: Equation to solve} becomes modified to the first-order differential equation
\begin{align}
    \left[\hat{H}^{(0)}, \hat{S}^{(j)}(t)\right] = \hat{P}^{(j)}(t) + i\hbar \frac{\partial \hat{S}^{(j)}(t)}{\partial t}. \label{eq: time dependent defining equation}
\end{align}
This equation captures the interplay between the transformation's goal, encoded in $\hat{P}^{(j)}(t)$, and the dynamical constraint imposed by the time evolution of the generator itself.

%%%%%%%%%%%%%%%%%%%%%%%%%%%%%%%%%%%%%%%%%%%%%%%%
%%%%%%%%%%%%%%%%%%%%%%%%%%%%%%%%%%%%%%%%%%%%%%%%
%%%%%%%%%%%%%%%%%%%%%%%%%%%%%%%%%%%%%%%%%%%%%%%%
\section{Universal Solution to the Generator}\label{sec: General solution to S generator}

The defining equation $[\hat{H}^{(0)}, \hat{S}^{(j)}] = \hat{P}^{(j)}$ stands at the heart of perturbative unitary transformations, yet as we have seen, solving it in practice has historically required either inspired guesswork for operator-level approaches or basis-dependent calculations that struggle with infinite-dimensional spaces. What we seek is a solution that transcends these limitations: one that works uniformly across a large variety of Hilbert space structures, preserves operator-level insights, requires no truncation, and applies regardless of which perturbative scheme we choose to implement. The key to achieving this universality lies in recognizing that quantum systems, despite their diversity, share a common mathematical structure in how operators act on their Hilbert spaces.

%%%%%%%%%%%%%%%%%%%%%%%%%%%%%%%%%%%%%%%%%%%%%%%%
\subsection{Operator Structure and Hilbert Space Decomposition}\label{sec: operator structure}

To develop our unified framework, we begin by characterizing the most general structure of operators and Hamiltonians within the types of Hilbert spaces commonly encountered in quantum physics. While our approach extends straightforwardly to systems with multiple bosonic modes or fermionic degrees of freedom (as detailed in Appendix~\ref{sec: appendix multiple bosonic subspaces}), we focus here on systems whose total Hilbert space decomposes as $\mathds{H} = \mathds{H}_f \otimes \mathds{H}_b$, where $\mathds{H}_f$ is a finite-dimensional subspace of dimension $d_f$ and $\mathds{H}_b$ is a bosonic Fock space characterized by the number operator $\hat{N} \equiv \hat{a}^\dagger \hat{a}$. This structure captures a remarkably broad class of physical systems: from cavity and circuit QED, where qubits couple to resonator modes, to quantum optomechanics, where mechanical oscillators interact with two-level systems, to the Jaynes-Cummings model and its variants that underpin much of quantum optics.

Within such a Hilbert space, any Hermitian operator $\hat{\mathcal{O}}^+$ or anti-Hermitian operator $\hat{\mathcal{O}}^-$ admits a canonical decomposition (see Appendix~\ref{sec: appendix proof of general operator} for the detailed derivation)
\begin{align}
    \hat{\mathcal{O}}^\pm = \sum_{\mu\nu = 0}^{d_f-1} \sum_{\Delta \geq 0} \hat{o}_{\mu\nu}^{(\Delta)}(\hat{N}) \hat{a}^{\Delta} \hat{\sigma}_{\mu\nu} \pm \text{h.c.}, \label{eq: general operator}
\end{align}
where $\hat{\sigma}_{\mu\nu} \equiv |\mu\rangle \langle\nu|$ represents the transition operator between eigenstates $\{|\mu\rangle\}$ of the finite-dimensional subspace, and $\hat{o}_{\mu\nu}^{(\Delta)}(\hat{N})$ denotes an arbitrary function of the number operator. The index $\Delta$ labels the change in boson number induced by the term, while the Hermitian conjugate ensures the appropriate symmetry.

To clarify this notation consider an unperturbed Hamiltonian $\hat{H}^{(0)}$. Since this operator must be diagonal in whatever basis we have chosen for our perturbative analysis, it takes the form
\begin{align}
    \hat{H}^{(0)} = \sum_{\mu = 0}^{d_f-1} \hat{f}_\mu(\hat{N}) \hat{\sigma}_{\mu\mu}, \label{eq: general unperturbed hamiltonian}
\end{align}
where each function $\hat{f}_\mu(\hat{N})$ encodes the energy landscape for the $\mu$-th level of the finite subspace as a function of the bosonic occupation number. This formulation encompasses purely separable harmonic systems where $\hat{f}_\mu(\hat{N}) = E_\mu + \hbar\omega \hat{N}$, anharmonic oscillators where higher-order terms in $\hat{N}$ appear, and even dispersively coupled systems where the energy of one subsystem depends on the state of another.

To illustrate the concrete application of this framework, consider the paradigmatic quantum Rabi model, described by the Hamiltonian $\hat{H}^{(0)} = \hbar\Omega_R \hat{N} + \frac{\hbar\Omega_Z}{2}\hat{\sigma}_z$, with two-level system (photon) frequency $\Omega_Z$ ($\Omega_R$), for the unperturbed part and the interaction $\hat{H}^{(1)} = \hbar g(\hat{a}^\dagger + \hat{a})\hat{\sigma}_x$ with strength $g$. Here the finite-dimensional subspace is spanned by the two-level system states $\{|\mu\rangle\}_{\mu \in \{0,1\}}$ corresponding to spin up and down. The functions $\hat{f}_\mu(\hat{N})$ that appear in Eq.~\eqref{eq: general unperturbed hamiltonian} are simply
\begin{align}
    \hat{f}_{0}(\hat{N}) = \hbar \Omega_R \hat{N} + \frac{\hbar\Omega_Z}{2}, \quad
    \hat{f}_{1}(\hat{N}) = \hbar \Omega_R \hat{N} - \frac{\hbar\Omega_Z}{2}.
    \label{eq:f01Rabi}
\end{align}
The interaction term $\hat{H}^{(1)} = \hbar g(\hat{a}^\dagger + \hat{a})\hat{\sigma}_x$ also fits naturally into the structure of Eq.~\eqref{eq: general operator}: it contains only terms with $\Delta = 1$ (single photon processes), couples between different spin states ($\mu \neq \nu$), and has coefficients $\hat{o}_{01}^{(1)}(\hat{N}) = \hat{o}_{10}^{(1)}(\hat{N}) = \hbar g$ that are independent of $\hat{N}$.

%%%%%%%%%%%%%%%%%%%%%%%%%%%%%%%%%%%%%%%%%%%%%%%%
\subsection{The Universal Solution}\label{sec: universal solution derivation}

With this operator structure established, we can now attack the defining equation $[\hat{H}^{(0)}, \hat{S}^{(j)}] = \hat{P}^{(j)}$ systematically. Both the target operator $\hat{P}^{(j)}$ and the generator $\hat{S}^{(j)}$ decompose according to Eq.~\eqref{eq: general operator}
\begin{align}
    \hat{P}^{(j)} &\equiv \sum_{\mu\nu = 0}^{d_f-1} \sum_{\Delta \geq 0} \hat{p}_{\mu\nu}^{(\Delta)}(\hat{N}) \hat{a}^{\Delta} \hat{\sigma}_{\mu\nu} + \text{h.c}, \label{eq: Pj}\\
    \hat{S}^{(j)} &\equiv \sum_{\mu\nu = 0}^{d_f-1} \sum_{\Delta \geq 0} \hat{s}_{\mu\nu}^{(\Delta)}(\hat{N}) \hat{a}^{\Delta} \hat{\sigma}_{\mu\nu} - \text{h.c},
\end{align}
where the coefficients $\hat{p}_{\mu\nu}^{(\Delta)}(\hat{N})$ and $\hat{s}_{\mu\nu}^{(\Delta)}(\hat{N})$ implicitly carry information about the perturbation order $j$.

The crucial algebraic step involves evaluating the commutator $[\hat{H}^{(0)}, \hat{S}^{(j)}]$. The finite-dimensional part yields straightforward results through the usual commutation relations of the transition matrices $\hat{\sigma}_{\mu\nu}$, but the bosonic part requires more care. The key identity is the operator relation $\hat{a}^\Delta \hat{f}(\hat{N}) = \hat{f}(\hat{N}+\Delta) \hat{a}^\Delta$, which captures how ladder operators shift the argument of number-dependent functions. This simple relation allows us to track how boson creation and annihilation operators thread through the energy landscape encoded in $\hat{f}_\mu(\hat{N})$.

Working through the commutator term by term and matching coefficients between the two sides of the defining equation yields the central result of this work
\begin{align}
    \hat{s}_{\mu\nu}^{(\Delta)}(\hat{N}) = -\frac{1}{\hbar\hat{\omega}_{\mu\nu}^{(\Delta)}} \hat{p}_{\mu\nu}^{(\Delta)}(\hat{N}),\label{eq: generator for time independent gamma}
\end{align}
where we have defined the energy difference operator
\begin{align}
    \hbar\hat{\omega}_{\mu\nu}^{(\Delta)} \equiv \hat{f}_\nu(\hat{N} + \Delta) - \hat{f}_\mu(\hat{N}). \label{eq: frequency operator}
\end{align}

This compact expression encodes the complete solution to our problem. Each coefficient $\hat{s}_{\mu\nu}^{(\Delta)}(\hat{N})$ of the generator is determined directly from the corresponding coefficient of the target operator $\hat{p}_{\mu\nu}^{(\Delta)}(\hat{N})$ through division by a transition frequency operator that captures the energy cost of the transition being induced (see Eq.~\eqref{eq: S1 for QRM} for an application of Eq.~\eqref{eq: generator for time independent gamma} on the quantum Rabi model example considered in this section). The operator $\hat{\omega}_{\mu\nu}^{(\Delta)}$ represents the difference in energy between the initial state (in level $\mu$ with $\hat{N}$ bosons) and the final state (in level $\nu$ with $\hat{N}+\Delta$ bosons), thereby encoding precisely the transition frequencies that govern the perturbative response of the system.

%%%%%%%%%%%%%%%%%%%%%%%%%%%%%%%%%%%%%%%%%%%%%%%%
\paragraph{Physical Interpretation and Connections:}\label{sec: physical interpretation}

The physical content of Eq.~\eqref{eq: generator for time independent gamma} deserves careful examination. In perturbation theory, we understand intuitively that weak perturbations mixing nearly degenerate states require more careful treatment than those connecting widely separated energy levels. This intuition manifests mathematically through the energy denominator structure familiar from Fermi's golden rule and standard perturbation theory. What Eq.~\eqref{eq: generator for time independent gamma} provides is the operator-level generalization of this: the frequency operator $\hbar \hat{\omega}_{\mu\nu}^{(\Delta)}$ plays exactly the role of the energy difference, but now it remains an operator acting on the Hilbert space rather than collapsing to a scalar. For systems without anharmonicities, where $\hat{f}_\mu(\hat{N}) = E_\mu + \hbar\omega \hat{N}$ for constants $E_\mu$ and $\omega$, the frequency operator simplifies to $\hbar \hat{\omega}_{\mu\nu}^{(\Delta)} = E_\nu - E_\mu + \hbar\omega\Delta$: precisely the transition frequency between states $|\mu, n\rangle$ and $|\nu, n+\Delta\rangle$.

This result unifies and extends several previous approaches to computing generators of perturbative transformations. The matrix-element formula in Eq.~\eqref{eq: sum over indices approach}, which required explicit summation over basis states, emerges immediately by taking matrix elements of Eq.~\eqref{eq: generator for time independent gamma} in the joint basis of finite-dimensional and Fock states. The operator frequency $\hbar \hat{\omega}_{\mu\nu}^{(\Delta)}$ reduces to the energy difference $E_\ell - E_h$ in this representation, recovering the traditional result but without requiring us to work exclusively in that basis. Similarly, the integral representation of Eq.~\eqref{eq: interaction picture solution} can be shown to yield Eq.~\eqref{eq: generator for time independent gamma}  (see Appendix~\ref{sec: appendix proof equivalence interaction} for details). Thus our closed-form solution provides the common foundation underlying these various perspectives.  

Recent work has emphasized the role of eigenoperators in understanding perturbative transformations. In Ref.~\cite{landi2024eigenoperatorapproachschriefferwolffperturbation}, the authors show that for harmonic systems, the generator can be constructed by identifying eigenoperators of the commutator superoperator $[\hat{H}^{(0)}, \cdot]$ and using their eigenvalues to determine the generator coefficients. The frequency operator $\hat{\omega}_{\mu\nu}^{(\Delta)}$ we have derived represents precisely these eigenvalues: when the system lacks anharmonicities, $\hat{\omega}_{\mu\nu}^{(\Delta)}$ equals the eigenvalue of the eigenoperator $\hat{p}_{\mu\nu}^{(\Delta)}(\hat{N}) \hat{a}^\Delta \hat{\sigma}_{\mu\nu}$ under the adjoint action of $\hat{H}^{(0)}$. Our framework thus provides a bridge between the eigenoperator perspective and the interaction picture approach of Ref.~\cite{SW_quantum_information_processing}, showing how both viewpoints emerge from the same underlying mathematical structure and extending their applicability to anharmonic systems where the simple eigenoperator picture must be generalized.

%%%%%%%%%%%%%%%%%%%%%%%%%%%%%%%%%%%%%%%%%%%%%%%%
\paragraph{Generality Beyond Standard Block Diagonalization:}\label{sec: beyond standard SWT}

It is crucial to emphasize that Eq.~\eqref{eq: generator for time independent gamma} applies to any perturbative unitary transformation scheme, not merely the standard SWT mentioned in Sec.~\ref{sec: perturbative unitary transformations}. The conventional SWT achieves block diagonalization between two subspaces (typically a low-energy manifold and a high-energy manifold) by eliminating their mutual coupling to a desired order in perturbation theory. This represents one particular choice for the target operators $\hat{P}^{(j)}$, but the structure of the defining equation $[\hat{H}^{(0)}, \hat{S}^{(j)}] = \hat{P}^{(j)}$ and its solution remain unchanged regardless of our physical objective.

When a system contains three or more distinct energy manifolds, the two-block SWT cannot achieve complete block diagonalization. Various modifications have been proposed to address this limitation: such as multi-block diagonalization schemes~\cite{SW_MultiblockDiagonalization, pymablock, DiVincenzo_multiblock} that systematically decouple multiple subspaces and recursive approaches~\cite{SW_RecursiveSW} that apply successive transformations to progressively eliminate off-diagonal structure. Each of these methods involves constructing appropriate defining equations for the generator $\hat{S}^{(j)}$ at each order, that is, specifying what $\hat{P}^{(j)}$ should be to achieve the desired simplification.

What makes our unified solution valuable is precisely that it remains agnostic to these choices. Once you have determined what transformation you want to implement, whether standard block diagonalization, multi-block decoupling, or some other reorganization of the Hamiltonian, the operator $\hat{P}^{(j)}$ encodes that choice, and Eq.~\eqref{eq: generator for time independent gamma} immediately provides the generator that implements it.

%%%%%%%%%%%%%%%%%%%%%%%%%%%%%%%%%%%%%%%%%%%%%%%%
%%%%%%%%%%%%%%%%%%%%%%%%%%%%%%%%%%%%%%%%%%%%%%%%
\subsection{Extension to Time-Dependent Systems}\label{sec: time dependence}

Having established a universal solution for static perturbative transformations, we now return to the challenge posed in Sec.~\ref{sec: time dependent transformations}: how does this framework extend when the Hamiltonian's perturbations carry explicit time dependence? The mathematical structure we have developed remains robust under this generalization, though the physical content becomes richer as we must now account for the dynamical effects carried by transformation's temporal evolution (see Appendix~\ref{appendix: sambe space} for an extension of our formalism to operators expressed in Sambe space). 

The operator decomposition of Eq.~\eqref{eq: general operator} remains valid for time-dependent systems, with the understanding that the coefficient functions now carry parametric time dependence. For a time-dependent perturbation, we write
\begin{align}
    \hat{P}^{(j)}(t) \equiv \sum_{\mu\nu = 0}^{d_f-1} \sum_{\Delta \geq 0} \hat{p}_{\mu\nu}^{(\Delta)}(\hat{N},t) \hat{a}^{\Delta} \hat{\sigma}_{\mu\nu} + \text{h.c},
\end{align}
and similarly for the generator,
\begin{align}
    \hat{S}^{(j)}(t) \equiv \sum_{\mu\nu = 0}^{d_f-1} \sum_{\Delta \geq 0} \hat{s}_{\mu\nu}^{(\Delta)}(\hat{N},t) \hat{a}^{\Delta} \hat{\sigma}_{\mu\nu} - \text{h.c}.
\end{align}
As discussed in Sec.~\ref{sec: time dependent transformations}, two distinct regimes emerge depending on whether the time derivative $\frac{\partial \hat{S}^{(j)}(t)}{\partial t}$ contributes at the same perturbative order as $\hat{S}^{(j)}(t)$ itself. In the adiabatic regime, where variations occur slowly compared to the system's internal dynamics, the static solution of Eq.~\eqref{eq: generator for time independent gamma} carries through with time-dependent coefficients
\begin{align}
    \hat{s}_{\mu\nu}^{(\Delta)}(\hat{N},t) = -\frac{1}{\hbar \hat{\omega}_{\mu\nu}^{(\Delta)}} \hat{p}_{\mu\nu}^{(\Delta)}(\hat{N},t). 
\end{align}
This result captures systems driven by slowly varying control fields or parametric modulation where the transformation can follow the perturbation quasi-statically.

The more intricate case arises in the non-adiabatic regime, where rapid modulation demands that we account for the time evolution of the generator. Recall from Sec.~\ref{sec: time dependent transformations} that the defining equation becomes %modified to
\begin{align}
    \left[\hat{H}^{(0)}, \hat{S}^{(j)}(t)\right] = \hat{P}^{(j)}(t) + i\hbar \frac{\partial \hat{S}^{(j)}(t)}{\partial t}.
\end{align}
Projecting this equation onto our operator basis by matching coefficients of $\hat{a}^\Delta \hat{\sigma}_{\mu\nu}$ on both sides, and utilizing the commutation properties that led to Eq.~\eqref{eq: generator for time independent gamma}, we find that each coefficient satisfies the first-order inhomogeneous differential equation
\begin{align}
    \hat{s}_{\mu\nu}^{(\Delta)}(\hat{N}, t) = \frac{-1}{\hbar \hat{\omega}_{\mu\nu}^{(\Delta)}} \left[\hat{p}_{\mu\nu}^{(\Delta)}(\hat{N}, t) + i\hbar \frac{\partial \hat{s}_{\mu\nu}^{(\Delta)}(\hat{N}, t)}{\partial t} \right].\label{eq: generator for time dependent gamma}
\end{align}
This equation shows how, for non-adiabatic drivings, the generator must balance the goal encoded in $\hat{p}_{\mu\nu}^{(\Delta)}(\hat{N},t)$ against the dynamical constraint imposed by its own rate of change, weighted by the transition frequency $\hat{\omega}_{\mu\nu}^{(\Delta)}$.

Among time-dependent perturbations, periodic driving occupies a privileged position both theoretically and experimentally. Microwave and radiofrequency control of superconducting circuits, optical lattices for ultracold atoms, parametric pumping in cavity systems, and AC gate voltages in semiconductor quantum dots all subject quantum systems to periodic modulation. When a perturbation repeats with period $T = 2\pi/\Omega$, the system's long-time behavior is governed by Floquet theory, the temporal analog of Bloch's theorem for spatial crystals. Perturbative unitary transformations provide the natural framework for computing the effective Floquet Hamiltonian that captures stroboscopic dynamics and quasi-energy spectra.

For a time-periodic perturbation with fundamental frequency $\Omega$, we expand each coefficient in its Fourier series. To maintain notational clarity, let us contract the indices $(\mu, \nu, \Delta)$ into a single index $k$, writing

\begin{align}
    \hat{p}_{\mu\nu}^{(\Delta)}(\hat{N},t) &\equiv \hat{p}_k(\hat{N},t) \\
    &=\sum_{n=-\infty}^\infty \hat{p}_k^n(\hat{N}) e^{in\Omega_k t},
\end{align}

where $\Omega_k$ is the fundamental frequency of the perturbation. By ensuring that Eq.~(\ref{eq: complete time dependent rotation}) preserves the system's macromotion~\footnote{Equation~\eqref{eq: generator for time dependent gamma} admits a homogeneous solution proportional to $e^{i\hat{\omega}_k t}$, which oscillates at the system's intrinsic transition frequency rather than the driving frequency. Preserving the macromotion amounts to setting this contribution to zero, leaving $\hat{S}^{(j)}(t)$ to describe only the drive-induced periodic micromotion.}~\cite{SW_TD_FloquetSW}, the solution to Eq.~(\ref{eq: generator for time dependent gamma}) is given by

\begin{align}
     \hat{s}_k(\hat{N}, t) =- \sum_{n=-\infty}^\infty \frac{\hat{p}_k^{(n)}(\hat{N}) e^{i n \Omega_k t}}{\hbar \hat{\omega}_k - n\hbar \Omega_k}. \label{eq: generator for periodic gamma}
\end{align}

The above equation constitutes the second main result of this paper. Similar to Eq.~(\ref{eq: generator for time independent gamma}), the computation of Eq.~(\ref{eq: generator for periodic gamma}) provides a systematic method for handling perturbations in time-periodically driven systems. Furthermore, with Eq.~(\ref{eq: generator for periodic gamma}) we gain additional insights regarding the treatment of perturbative time dependent interactions. For time independent perturbative transformations the only requirement is to ensure the perturbation to be much smaller than the energy difference of the coupled states (see Eq.~(\ref{eq: sum over indices approach})). However, Eq.~(\ref{eq: generator for periodic gamma}) indicates that for periodic perturbations, whose Fourier series contains no static component, the above mentioned requirement changes to $||\hat{P}^{(n)}_k||\ll |\hbar \hat{\omega}_k - n \hbar\Omega_k|$ for $n\in \mathds N_{\geq 0}$.

\section{Examples}

\subsection{Dispersive shift in anharmonic systems} \label{sec: example 1 system}

The interaction between quantum two-level systems and resonators forms one of the many cornerstones of modern quantum information processing.  To demonstrate the power and practical utility of our framework, we analyze  the paradigmatic Rabi model considered in Sec.~\ref{sec: universal solution derivation} with the photonic mode rendered anharmonic. The system is described by the Hamiltonian
\begin{align}
&\hat H=\hat H^{(0)}+\lambda\hat H^{(1)},\\
%%%%%%%%%%%%%%%%%%%%%%%%%%%%%%%%%%%%%%%%%%%%%
&\hat{H}^{(0)} = \hbar \Omega_R \hat{N}+ \hbar\alpha \hat{N}^2 + \frac{\hbar \Omega_Z}{2} \hat{\sigma}_z, \label{eq: toy model hamiltonian H0}\\
%%%%%%%%%%%%%%%%%%%%%%%%%%%%%%%%%%%%%%%%%%%%%
&\hat{H}^{(1)} = \hbar g\left(\hat{a} + \hat{a}^\dagger\right)\hat{\sigma}_x, \label{eq: toy model hamiltonian V}
\end{align}
where the unperturbed contribution captures the bare TLS frequency $\Omega_Z$ and anharmonic oscillator with frequency $\Omega_R$ anharmonicity parameter $\alpha$, while the perturbative coupling represents the transverse light-matter interaction, with strength $g$, that hybridizes the TLS and oscillator degrees of freedom.

While the idealized harmonic oscillator provides a tractable starting point for analysis, realistic implementations invariably exhibit anharmonicity: whether from the Josephson nonlinearity in transmon qubits, Kerr effects in optical cavities, or intrinsic material properties in mechanical resonators. The anharmonic contribution $\alpha\hat N^2$ in Eq.~\eqref{eq: toy model hamiltonian H0} introduces a systematic occupation-number dependence in the oscillator's level spacing, fundamentally altering the system's response to the coupling term in Eq.~\eqref{eq: toy model hamiltonian V}.

Traditional perturbative approaches to such systems face a methodological dilemma when confronting the anharmonic term. The anharmonicity represents a crucial physical effect that strongly influences system dynamics, particularly at higher photon numbers where dispersive readout and multi-photon processes become more appreciable. Yet conventional perturbative approaches often relegate anharmonicity to the perturbative sector alongside the coupling term~\cite{SW_beyond_dispersive_regime_Ansatz,landi2024eigenoperatorapproachschriefferwolffperturbation, SW_quantum_information_processing,SW_I_Example} (i.e. setting $\alpha\propto\lambda$, rather than $\alpha\propto \lambda^0$).

The framework developed in this work eliminates this artificial constraint. The unperturbed Hamiltonian in Eq.~\eqref{eq: toy model hamiltonian H0} admits a transparent decomposition within our general framework. The finite-dimensional subspace is spanned by the TLS eigenstates $\{|0\rangle, |1\rangle\}$ of $\hat{\sigma}_z$, while the infinite-dimensional bosonic sector is characterized by the Fock states $\{|n\rangle\}$ of the anharmonic oscillator. Following Eq.~\eqref{eq: general unperturbed hamiltonian}, we identify the number-dependent energy functions
\begin{align}
\hat{f}_{\nu}(\hat{N}) &= \hbar  \Omega_R \hat{N} + \hbar\alpha \hat{N}^2 +(-1)^\nu \frac{\hbar \Omega_Z}{2}, \label{eq: f0 toy model}
\end{align}
which encode how the total system energy depends on both the TLS state (indexed by $\nu \in \{0,1\}$) and the oscillator occupation number $\hat{N}$. 

The perturbation in Eq.~\eqref{eq: toy model hamiltonian V} also fits naturally into the operator decomposition of Eq.~\eqref{eq: general operator}. Expanding $\hat{\sigma}_x = \hat{\sigma}_{01} + \hat{\sigma}_{10}$ and noting that the interaction changes the bosonic occupation by $\Delta = 1$ (single-photon processes), we identify the coefficient functions
\begin{align}
\hat{p}^{(1)}_{01}(\hat{N}) = \hat{p}^{(1)}_{10}(\hat{N}) = -\hbar g, \label{eq: g for toy model}
\end{align}
with all other components vanishing. The interaction thus couples states that differ both in TLS index and photon number, driving transitions between $|\mu, n\rangle \leftrightarrow |\nu, n\pm1\rangle$ for $\mu \neq \nu$.

To construct the effective Hamiltonian through second order in perturbation theory, we aim at eliminating the off-diagonal TLS-oscillator coupling. Setting $\hat{P}^{(1)} = -\hat{H}^{(1)}$ in accordance with Eq.~\eqref{eq: equation to solve s1}, the generator $\hat{S}^{(1)}$ must satisfy $[\hat{H}^{(0)}, \hat{S}^{(1)}] = -\hat{H}^{(1)}$. Our universal solution in Eq.~\eqref{eq: generator for time independent gamma} immediately provides the generator coefficients. For the non-zero components corresponding to single-photon processes coupling different TLS states ($\mu\neq \nu$), we obtain
\begin{align}
    \hat s^{(1)}_{\mu\nu}(\hat N) \equiv \frac{g}{\hat \omega_{\mu\nu}^{(1)}}.\label{eq: s1_munu}
\end{align}
where the frequency operators, computed via Eq.~\eqref{eq: frequency operator}, encode the energy cost of each transition:
\begin{align}
\hbar\hat{\omega}^{(1)}_{\mu\nu} &= \hat{f}_\nu(\hat{N}+1) - \hat{f}_\mu(\hat{N}) \\
&= \hbar\left[\Omega_R + \alpha(2\hat{N}+1) + (-1)^\nu \Omega_Z\right], \label{eq: omega plus toy}
\end{align}
for $\mu \neq \nu$. These expressions reveal how anharmonicity modifies the transition frequencies: whereas a harmonic oscillator would yield constant denominators $\Omega_R \pm \Omega_Z$, the anharmonic correction introduces an explicit dependence on photon number through the $2\alpha\hat{N}$ term. This number-dependent detuning fundamentally alters the strength of virtual processes at different occupation levels, an effect that cannot be captured when anharmonicity is treated perturbatively (i.e. setting $\alpha\propto\lambda$).

The second-order effective Hamiltonian follows from the standard Baker-Campbell-Hausdorff expansion, retaining terms through second order in $\lambda$
\begin{align}
\hat{\mathcal{H}} = \hat{H}^{(0)} + \frac{\lambda^2}{2}\left[\hat{H}^{(1)}, \hat{S}^{(1)}\right]. \label{eq: Heff second order}
\end{align}
The commutator generates effective interactions between states within the same TLS subspace, mediated by virtual excitations to the opposite TLS state. Explicitly evaluating this commutator using the generator coefficients from Eq.~\eqref{eq: s1_munu} and organizing terms according to their operator structure, we obtain
\begin{align}
    \hat{\mathcal{H}} = \hat H^{(0)} + \hbar\lambda^2\left[\hat \delta(\hat N) + \hat \xi(\hat N)\hat\sigma_z \right], \label{eq: effective hamiltonian complete}
\end{align}
with
\begin{widetext}
    \begin{align}
        &\hat \delta(\hat N) \equiv \frac{g^2}{2}\left[\frac{\hat N+1}{\Omega_- - \alpha\left( 2\hat N + 1\right)} - \frac{\hat N+1}{\Omega_+ + \alpha\left( 2\hat N+1\right)} -\frac{\hat N}{\Omega_--\alpha\left(2\hat N-1 \right)}+\frac{\hat N}{\Omega_++\alpha\left(2\hat N-1 \right)}\right],\\
        %%%%%%%%%%%%%%%%%%%%%%%%%%%%%%%%%%%%%%%%%%%%%%%%%%%%%%%%%%%%%%%55
        &\hat \xi(\hat N) \equiv \frac{g^2}{2}\left[\frac{\hat N+1}{\Omega_- - \alpha\left( 2\hat N + 1\right)} 
        + \frac{\hat N+1}{\Omega_+ + \alpha\left( 2\hat N+1\right)} 
        +\frac{\hat N}{\Omega_--\alpha\left(2\hat N-1 \right)}+\frac{\hat N}{\Omega_++\alpha\left(2\hat N-1 \right)} \right], \label{eq: xi(N)}
    \end{align}
\end{widetext}
being functions of the number operator and $\Omega_\pm \equiv \Omega_Z \pm \Omega_R$. Note that in Eq.~\eqref{eq: effective hamiltonian complete} we omitted the off-diagonal two-photon process terms arising from the commutator in Eq.~\eqref{eq: Heff second order} as these would be eliminated by a second-order generator term $\hat S^{(2)}$.

The first term in Eq.~\eqref{eq: effective hamiltonian complete}, represents the zeroth order contribution, Eq.\eqref{eq: toy model hamiltonian H0}. The second term  provides a second order correction to both the bare resonator gap as well as to its anharmonicity and higher order non linearities. Lastly, the third term, proportional to $ \hat{\sigma}_z$, captures the dispersive interaction: the spectrum of the oscillator depends on the TLS state and, reciprocally, the TLS splitting depends on photon number. This cross-coupling enables the non-destructive readout that underpins dispersive qubit measurements.

The physical significance of including anharmonicity in $\hat{H}^{(0)}$ rather than treating it perturbatively becomes apparent upon examining the anharmonicity dependence of the frequency operators in Eq.~\eqref{eq: omega plus toy}
and the  effective coupling strength. 
%scales inversely with products of transition frequencies, and these frequencies themselves depend on $\hat{N}$ through the anharmonic term $\alpha(2\hat{N} \pm 1)$. 
In standard approaches in which the 
 anharmonicity is relegated to the perturbation, 
 %standard approaches must approximate these denominators 
the frequency operators are approximated 
 by their harmonic values
 and the anharmonic corrections are incorporated 
% , then attempt to incorporate anharmonic corrections 
through higher-order terms (see Appendix~\ref{appendix: perturbative anharmonicity}). Our approach, by contrast, incorporates the full $\hat{N}$-dependence of transition frequencies from the outset, automatically capturing how anharmonicity modulates the strength of virtual processes across the entire Fock space.

\begin{figure*}[tp]
    \centering
    \subfloat[]{
        \includegraphics[width=0.33\linewidth]{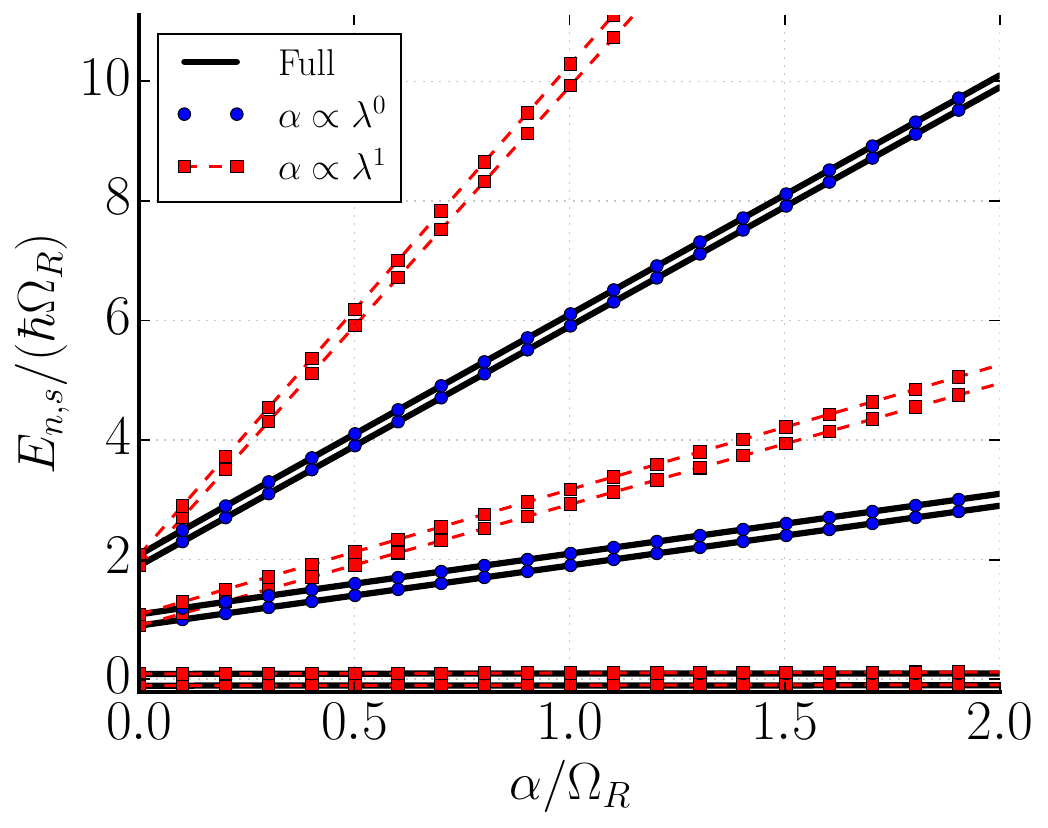}
        \label{fig:ex1_energy}
    }
   % \hfill
    \subfloat[]{
        \includegraphics[width=0.33\linewidth]{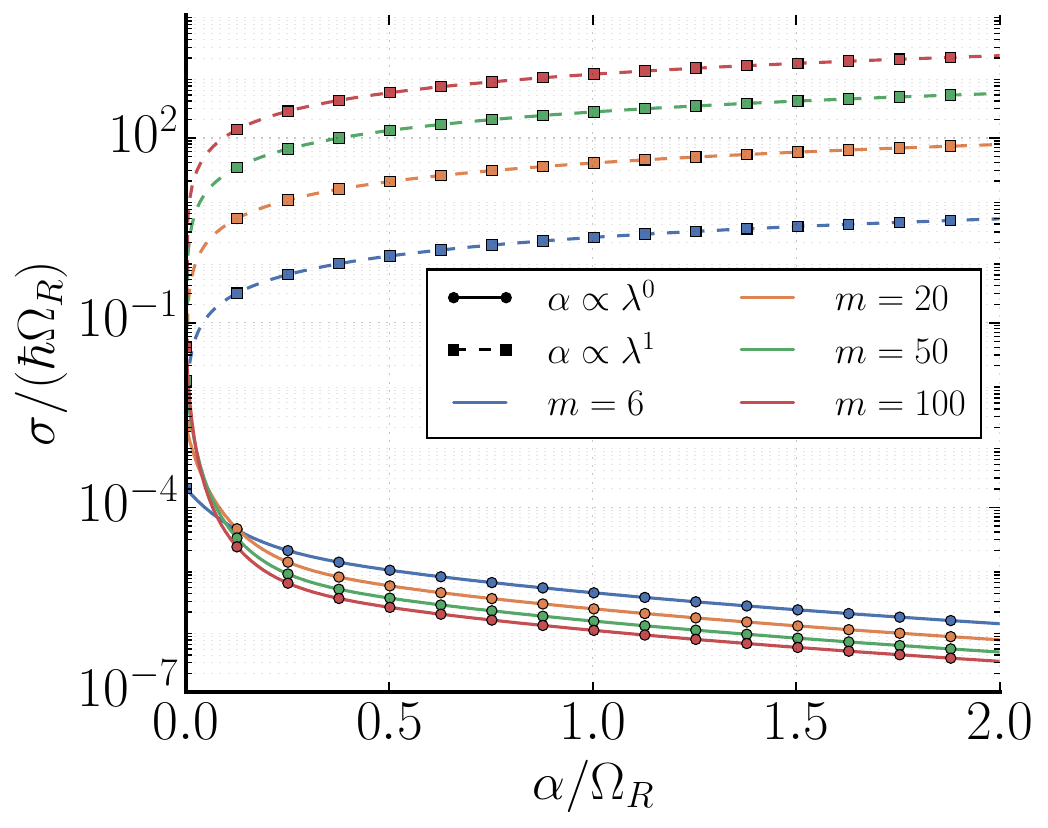}
        \label{fig:ex1_rms}
    }
 %   \hfill
    \subfloat[]{
        \includegraphics[width=0.33\linewidth]{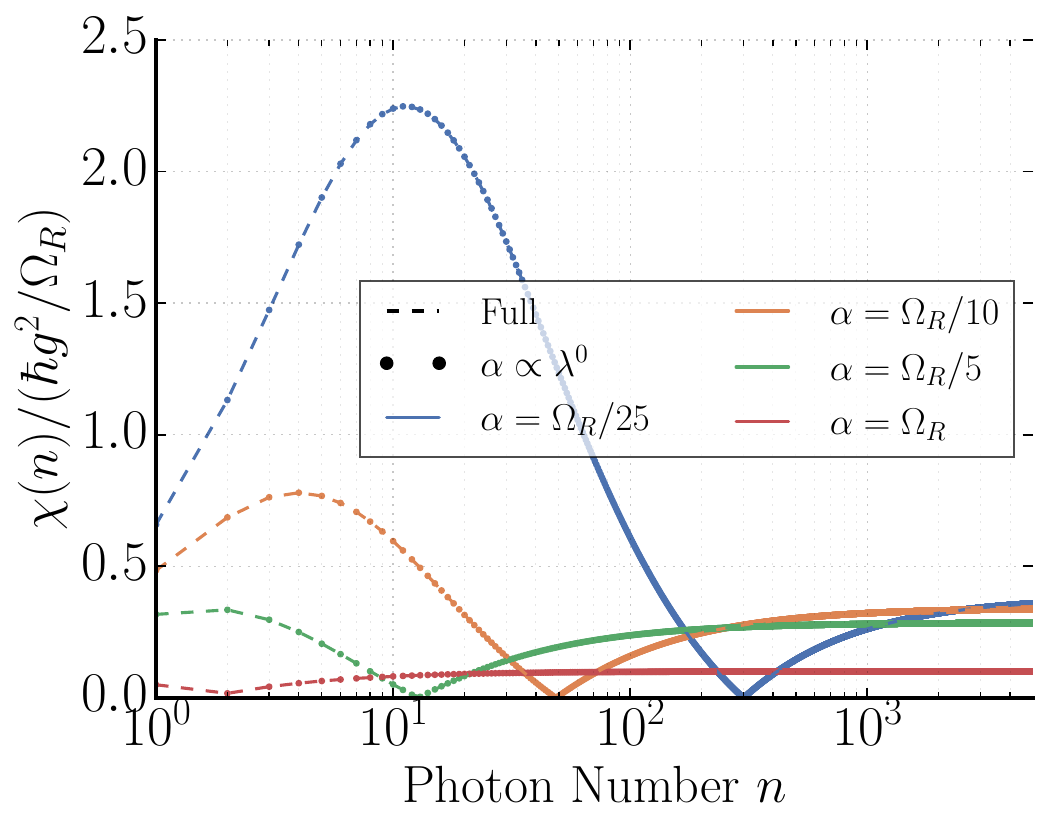}
        \label{fig:ex1_chi}
    }
    \caption{(a) Energy spectra from the full Hamiltonian (numerical), the second-order effective Hamiltonian $\mathcal{H}_\mathrm{eff}$ (blue), and the perturbative treatment of the anharmonicity (red) (see Appendix~\ref{appendix: perturbative anharmonicity}). (b) RMS error $\sigma/(\hbar\Omega_R)$ of the first $m$ energy levels, where the perturbative approach (see Appendix~\ref{appendix: perturbative anharmonicity}) (dashed line with squares) deviates increasingly with $\alpha$ and number of levels, while our method (solid line with circles) remains accurate and improves with larger $\alpha$. (c) Dispersive shift as a function of photon number. The simulation parameters are $\Omega_Z = 0.2\,\Omega_R$ and $g = 0.1\,\Omega_R$.}
    \label{fig:example1}
\end{figure*}

To quantitatively assess the accuracy of this improved treatment, we compare the energy spectrum predicted by our diagonal $\hat{\mathcal{H}}$, Eq.~\eqref{eq: effective hamiltonian complete}, with numerical diagonalization of the full Hamiltonian in Eqs.~\eqref{eq: toy model hamiltonian H0}--\eqref{eq: toy model hamiltonian V}. Figure~\ref{fig:ex1_energy} displays the lower-lying energy levels for representative parameters (see figure caption). The energy levels $E_{n,0(1)}$ obtained from our effective Hamiltonian (blue dots) track the exact numerical results (black lines) with remarkable fidelity across both TLS states and all displayed photon numbers. In stark contrast, the perturbative treatment of anharmonicity (dashed red curves), which attempts to account for the $\alpha\hat N^2$ term as a first-order correction to an otherwise harmonic system, exhibits systematic deviations for $n>0$ that grow progressively worse at higher photon numbers.

This qualitative observation finds quantitative support in the root-mean-square (RMS) error analysis presented in Fig.~\ref{fig:ex1_rms}. For a given anharmonicity $\alpha$, we compute the RMS deviation between predicted and exact energies for the lowest $m$ levels, normalized by the characteristic energy scale $\hbar\Omega_R$. The perturbative treatment (dashed curves with square markers) displays two troubling features: first, the error grows monotonically with increasing anharmonicity, reaching several percent of $\hbar\Omega_R$ even for modest values $\alpha/\Omega_R \sim 0.1$; second, including more energy levels in the RMS calculation systematically worsens the error, indicating that the approximation deteriorates at higher photon numbers where nonlinearity becomes pronounced. Our method (solid curves with circular markers) exhibits the opposite behavior on both counts: the accuracy improves as anharmonicity increases (reflecting that larger $\alpha$ enhances the separation between the perturbative coupling and the anharmonic energy scale), and incorporating additional levels in the RMS average reduces the error, demonstrating convergent behavior consistent with a well-controlled perturbative expansion.

The dispersive shift $\chi(n)$, defined as the TLS-state-dependent change in oscillator transition frequency, provides a physically transparent metric for evaluating these competing approaches. For a state $|n,s\rangle$ where $n$ denotes photon number and $s \in \{0,1\}$ labels the TLS state, the transition frequency from $|0,s\rangle$ to $|n,s\rangle$ is determined by the energy difference $E_{n,s} - E_{0,s}$. The dispersive shift quantifies how much this frequency differs between the two TLS states
\begin{align}
\chi(n) \equiv \left| \left(E_{n,0} - E_{0,0} \right) - \left(E_{n,1} - E_{0,1} \right)\right|. \label{eq: dispersive shift definition}
\end{align}
In circuit QED readout protocols, the oscillator is set at a frequency chosen to maximize the separation between responses conditioned on the TLS state; a larger dispersive shift $\chi(n)$ translates directly to improved measurement fidelity and reduced integration time \cite{Dispersive_shift, Dsipersive_shift_2, Chi_optimize_readout}.
According to our  effective Hamiltonian in Eq.~\eqref{eq: effective hamiltonian complete}, the dispersive shift reads
\begin{align}
   \chi(n)
=
2\hbar
\left|
\xi(n)-\xi(0)
\right| \label{eq: chi(n)},
\end{align}
which is shown in Figure~\ref{fig:ex1_chi} 
%showcases the quantity derived in Eq.~\eqref{eq: chi(n)} 
as a function of the photon number $n$. To understand the underlying structures displayed in this plot, we refer to Eq.~\eqref{eq: xi(N)}. For positive $\alpha$ and $\Omega_R>\Omega_Z$, the first maximum of $\chi(n)$ occurs while $\xi(n)-\xi(0) <0$ and is therefore set by the minimum of $\xi(n)$. The complete expression for this is algebraically involved, however, its series expansion around $\alpha/\Omega_R\approx0$ can be found to be 
\begin{align}
    n_{\rm max} \approx \frac{\sqrt{\Omega_R^2-\Omega_Z^2}}{2\alpha}
    -
    \frac{1}{2}
    \left(
        1+
        \frac{\Omega_R}
        {\sqrt{\Omega_R^2-\Omega_Z^2}}
    \right)
    \label{eq: example1 maximum position}
\end{align}

Beyond the maxima of $\chi(n)$, the increasing anharmonic detuning causes $\chi(n)$ to return towards zero. Since $\xi(0)$ remains finite and negative for the parameters considered here, we require $\xi(n)$ to go below the value of $\xi(0)$ for $\xi(n) - \xi(0)$ to cross the zero line. This reflects into the dip of $\chi(n)$  towards the zero line shown in Fig.~\ref{fig:ex1_chi}. The position of this zero is found as a solution of $\xi(n_{\rm dip})=\xi(0)$, and by expansion around $\alpha/\Omega_R\approx0$ can be found to be
\begin{align}
    n_{\rm dip} \approx \frac{1}{2}+\frac{\Omega_R^2-\Omega_Z^2}{2\alpha^2}   +
    \frac{\Omega_R\alpha}
    {\Omega_R^2-\Omega_Z^2} \label{eq: example1 dip position}.
\end{align}
The position of the dips consequently moves as $\alpha^{-2}$, which is parametrically faster than the maxima positions found in Eq.~\eqref{eq: example1 maximum position}, which instead move only as $\alpha^{-1}$. The distinct scalings of $n_{\rm max}$ and $n_{\rm dip}$ imply that the interval between the maximum and the zero crossing varies as $\alpha$ is swept, thus making the anharmonicity a direct control knob for the width of the nonmonotonic dispersive regime.

The behavior following the zero crossings is instead governed by the suppression of the virtual transitions due to large occupation. As $n \to \infty$, the anharmonic terms $\alpha(2n \pm 1)$ dominate the frequency operators, yielding
\begin{align}
\lim_{n \to \infty} \chi(n) = \frac{2g^2\Omega_Z}{\hbar\left[\left(\Omega_R + \alpha\right)^2 - \Omega_Z^2\right]}, \label{eq: chi infinity}
\end{align}

representing the dispersive shift in the high-occupation regime.

%%%%%%%%%%%%%%%%%%%%%%%%%%%%%%%%%%%%%%%%%%%%%%%%
%%%%%%%%%%%%%%%%%%%%%%%%%%%%%%%%%%%%%%%%%%%%%%%%
%%%%%%%%%%%%%%%%%%%%%%%%%%%%%%%%%%%%%%%%%%%%%%%%
%%%%%%%%%%%%%%%%%%%%%%%%%%%%%%%%%%%%%%%%%%%%%%%%
%%%%%%%%%%%%%%%%%%%%%%%%%%%%%%%%%%%%%%%%%%%%%%%%
%%%%%%%%%%%%%%%%%%%%%%%%%%%%%%%%%%%%%%%%%%%%%%%%
%%%%%%%%%%%%%%%%%%%%%%%%%%%%%%%%%%%%%%%%%%%%%%%%
%%%%%%%%%%%%%%%%%%%%%%%%%%%%%%%%%%%%%%%%%%%%%%%%

\subsection{Floquet dispersive shift}~\label{sec: time dependent example}
%In this section we explore how our formalism applies to 
To showcase the use of out framework for
periodically driven systems, 
%Towards that aim, 
we consider a driven version of the quantum Rabi model. The system is described by the time-periodic Hamiltonian
\begin{align}
    \hat{H}(t) &= \hat{H}^{(0)}  + \lambda\hat H^{(1)}(t), \label{eq: driven rabi hamiltonian}\\
    %%%%%%%%%%%%%%%%%%%%%%%%%%%%%%%%%%%
    \hat{H}^{(0)} &= \hbar\Omega_R \hat{a}^\dagger \hat{a}+\frac{\hbar\Omega_Z}{2}\hat{\sigma}_z, \label{eq: driven rabi H0}\\
    %%%%%%%%%%%%%%%%%%%%%%%%%%%%%%%%%
    \hat{H}^{(1)}(t) &= \hbar g\cos(\Omega t)\left(\hat{a}^\dagger+\hat{a}\right)\hat{\sigma}_x . \label{eq: driven rabi Vt}
\end{align}
Here $\Omega_R$ denotes the oscillator frequency, $\hbar \Omega_Z$ the two-level-system splitting and $g$ the amplitude of the perturbative coupling rate, which is periodically modulated in time with driving frequency $\Omega$. Typical analytical treatments of this, or otherwise similar, time-dependent systems require limiting the driving frequency to a specific regime in which the problem becomes more tractable. This is the case of the high-frequency expansion often employed in Floquet systems to simplify the dynamics by averaging over the rapid micromotion and organizing the effective Hamiltonian as an expansion in inverse powers of $\Omega$~\cite{SW_TD_Floquet_quantum_to_classical_crossover,SW_TD_FloquetSW}. Such an approach is well controlled only when the driving frequency is the largest relevant energy scale of the problem. It therefore provides limited information in the intermediate-frequency regime. Conversely, adiabatic treatements assume that the drive varies slowly compared with the internal dynamics and are consequently restricted to the opposite limit~\cite{SW_TD_berry_phase_induced_entanglement}. In this section we showcase how the implementation of the formalism presented in Sec.~\ref{sec: General solution to S generator} allows to   obtain general  results in between these limiting cases (see Appendix~\ref{appendix: edge frequency example} for a similar analysis performed in the high and low driving frequency limits). 

%For the model in Eqs.~\eqref{eq: driven rabi hamiltonian}--\eqref{eq: driven rabi Vt}, the finite-dimensional subsystem is spanned by the eigenstates of $\hat \sigma_z$, while the bosonic sector is described by the number operator $\hat N=\hat a^\dagger\hat a$. 
In the notation introduced in Eq.~\eqref{eq: general unperturbed hamiltonian}, the unperturbed Hamiltonian is specified by
\begin{align}
    \hat{f}_{0}(\hat{N}) = \hbar\Omega_R \hat{N}+\frac{\hbar\Omega_Z}{2},
    \quad
    \hat{f}_{1}(\hat{N}) = \hbar\Omega_R \hat{N}-\frac{\hbar\Omega_Z}{2}. \label{eq: driven rabi f1}
\end{align}
The perturbation changes the oscillator occupation by one quantum and simultaneously flips the two-level system. Using the notation introduced with Eq.~\eqref{eq: general operator} and writing $\hat \sigma_x\equiv\hat\sigma_{01} + \hat \sigma_{10}$, its only non-vanishing coefficient functions are
\begin{align}
    \hat o_{01}^{(1)}(t)=\hat o_{10}^{(1)}(t) = \hbar g \cos(\Omega t),
\end{align}

In the following we perform a  perturbative transformation aimed at eliminating $\hat H^{(1)}(t)$, and thus $\hat P^{(1)}(t)\equiv -\hat H^{(1)}(t)$, leading to $\hat{p}_{\mu\nu}^{(1)}(t)=-\hat o_{\mu\nu}^{(1)}(t)$. According to Eq.~\eqref{eq: generator for periodic gamma}, the appropriate generator reads
\begin{align}
    &\hat S^{(1)}(t) = \sum_{\mu,\nu=0,1,\mu\neq\nu}\hat{s}^{(1)}_{\mu\nu}(t)\hat a\hat \sigma_{\mu\nu} - \text{h.c.},\label{eq: S1 time dependent example low freq}\\
    %%%%%%%%%%%%%%%%%%%%%%%%%%%%%%%%%%%%%
    &\hat s^{(1)}_{01}(t) = -\frac{g}{2}\left[\frac{e^{i\Omega t}}{\Omega_-+\Omega} + \frac{e^{-i\Omega t}}{\Omega_--\Omega} \right],\\
    %%%%%%%%%%%%%%%%%%%%%%%%%%%%%%%%%%%%%
    &\hat s^{(1)}_{10}(t) = \frac{g}{2}\left[\frac{e^{i\Omega t}}{\Omega_+-\Omega} + \frac{e^{-i\Omega t}}{\Omega_++\Omega} \right].
\end{align}
with $\Omega_\pm \equiv \Omega_Z \pm \Omega_R$. 
%Using Eq.~\eqref{eq: S1 time dependent example low freq}, the effective Hamiltonian is then evaluated perturbatively to arrive at
From Eq.~\eqref{eq: Heff second order}, we then find the diagonal second-order effective Hamiltonian
\begin{widetext}
\begin{align}
    \hat{\mathcal H}(t) %=& \hat H^{(0)} + \frac{\lambda^2}{2}\left[\hat H^{(1)}(t), \hat S^{(1)}(t) \right] \\
    %%%%%%%%%%%%%%%%%%%%%%
    =&\hat H^{(0)} + \lambda^2\frac{\hbar g^2}{2} \left( \frac{\Omega_-}{\Omega_-^2-\Omega^2} + \frac{\Omega_+}{\Omega_+^2-\Omega^2}\right)\cos^2(\Omega t)\left(2\hat N + 1\right)\hat \sigma_z,~\label{eq: TD effective hamiltonian complete}
\end{align}
\end{widetext}
where, as in Sec.~\ref{sec: example 1 system}, we neglected two-photon processes, as these would be accounted by the next order generator $\hat S^{(2)}$. 
%Note that Eq.~\eqref{eq: TD effective hamiltonian complete} still retains a time dependence after the perturbative transformation. 
%However, in stark comparison to its parent Hamiltonian Eq.~\eqref{eq: driven rabi hamiltonian}, the newly obtained effective Hamiltonian 

Although this effective Hamiltonian is time dependent, it
commutes with itself at all times. 
%This implies that, in this perturbative frame, it is possible to obtain closed form 
therefore we can obtain  analytical expression for its Floquet quasi-energy operator integrating over one period 
%of the system, 
%without requiring a high frequency approximation; a often a common prescription for time dependent systems.
%
%With this insight, the quasi energies are obtained by evaluating the eigenphases accumulated by the system over one driving period $T=2\pi/\Omega$ 
(note we set $\lambda=1$),
\begin{align}
    \hat{\mathcal H}_{F}
&=\frac{1}{T}
\int_0^T
\hat{\mathcal H}(t)dt\\
%%%%%%%%%%%%%%%%%%%%%%%%%%%%%%%%%%%
&=\hat H^{(0)} + \hbar \frac{g^2}{2}\left[ \frac{\Omega_-}{\Omega_-^2 - \Omega^2} + \frac{\Omega_+}{\Omega_+^2 - \Omega^2}\right]\left(\hat N +\frac{1}{2}\right)\hat \sigma_z.
\label{eq: TD one period evolution}
\end{align}
One can then define a Floquet dispersive interaction that can be controlled via the driving frequency.
%provides an additional control parameter for tuning the dispersive interaction. 
% Additionally, we note how in Eq.~\eqref{eq: TD one period evolution}, the driving frequency $\Omega$ plays a similar role to the $\propto\alpha$ terms appearing in the static anharmonic case in Eq.~\eqref{eq: xi(N)}. This will become most apparent in proximity of $\Omega\approx\Omega_\pm$, for which we find diverging points due to resonances with the bare system's splittings.
% %, that equivalently occur also at analog positions in the static case.
%
Analogously to the definition of dispersive shift in Eq.~\eqref{eq: dispersive shift definition}, 
we define the Floquet dispersive shift using the quasienergies in this case. This quantity,
%for the static example, we here introduce the notion of dispersive shift in terms of the phase accumulated by the eigenstates of the system over a period of oscillations 
\begin{align}
    \chi_F(\Omega, n)\equiv\hbar g^2 \left|\frac{\Omega_-}{\Omega_-^2 - \Omega^2} + \frac{\Omega_+}{\Omega_+^2 - \Omega^2}\right|n\ ,\label{eq:chi_f}
\end{align}
%This quantity 
measures the spin-dependent modification of the oscillator quasienergy spacing between the $0$ and $n$ photonic states.
%and therefore the periodically driven analogue of the static dispersive shift discussed in Sec.~\ref{sec: example 1 system}.
%
The low- and high-frequency limiting cases of this expression read
\begin{align}
    \chi_F^{\rm low} =    2\hbar g^2
    \left|
        \frac{\Omega_Z}{\Omega_Z^2-\Omega_R^2}
    \right|n,
    \quad
    \chi_F^{\rm high} = 2\hbar g^2
    \left|
        \frac{\Omega_Z}{\Omega^2}
    \right|n \label{eq: flowquet edge dispersive shifts}
\end{align}
which coincide with the ones obtained using standard approximate methods (see Appendix~\ref{appendix: edge frequency example}).

\begin{figure}[tp]
    \centering
    \includegraphics[width=1\linewidth]{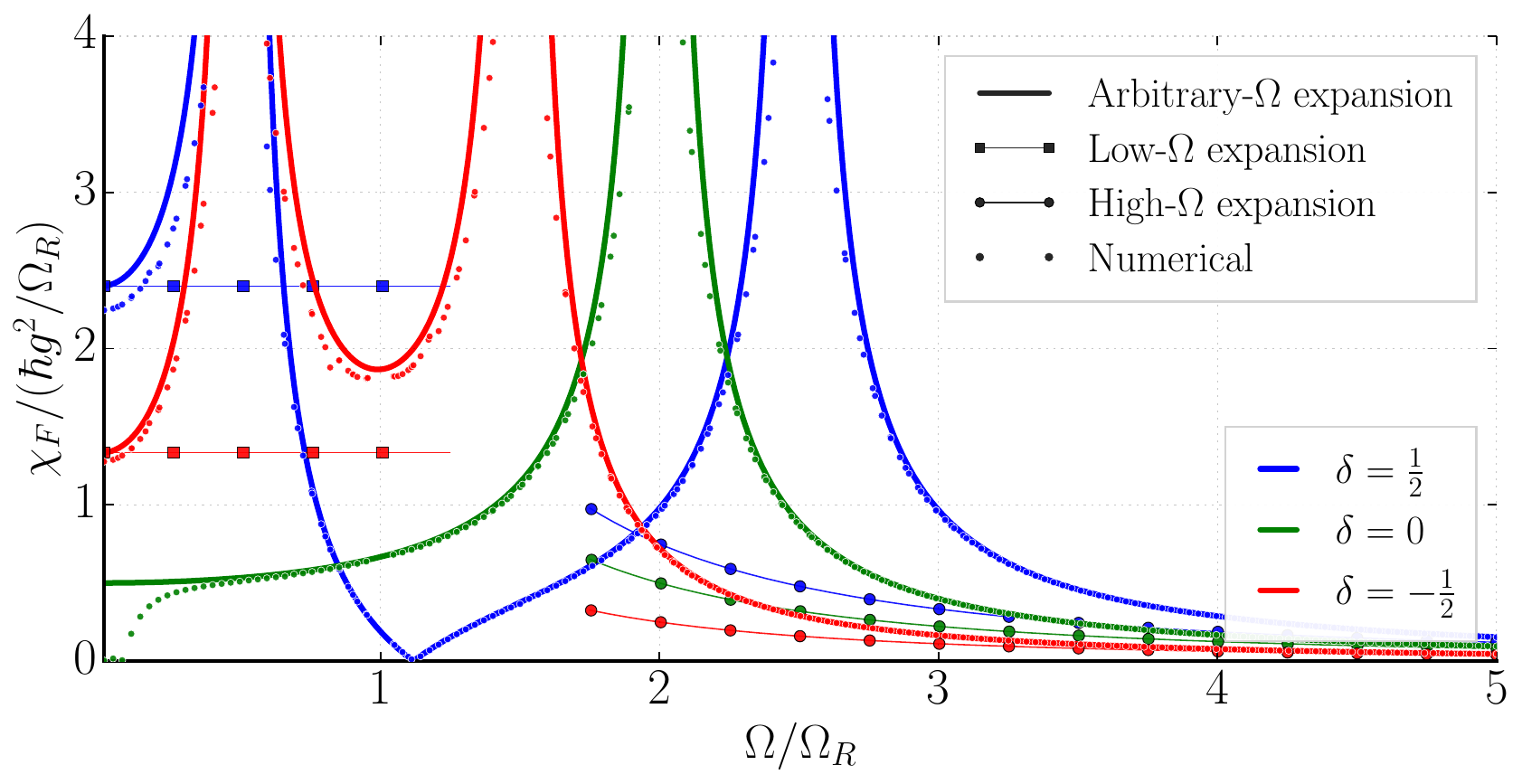}
        \caption{Floquet dispersive shift $\chi_F(\Omega, n=1)$ as a function of the normalized driving frequency $\Omega/\Omega_R$ for $\Omega_Z=\Omega_R (1 + \delta)$ where $\delta = \frac12$ (blue), $\delta=0$ (green), and $\delta=-\frac12$ (red), while the coupling strength is $g=0.1\Omega_R$. Solid lines show the arbitrary-frequency analytical second-order result obtained from Eq.~\eqref{eq:chi_f}, whereas the squares and dotted solid lines show the low- and high-frequency expansions in Eqs.~\eqref{eq: flowquet edge dispersive shifts}.
        The low-frequency limiting expression for $\Omega_R = \Omega_Z$ is not shown because $\Omega_- = 0$ and Eq. \eqref{eq: low frequency chi F} would diverge.
        The dots are obtained from a numerical Floquet calculation based on the one-period evolution operator, with the quasienergy branches tracked continuously as a function of $\Omega$.}
    \label{fig:chi_F}
\end{figure}

Figure~\ref{fig:chi_F} compares the arbitrary-frequency analytical result for the Floquet dispersive shift, Eq.~\eqref{eq:chi_f}, with the low- and high-frequency expressions in Eqs.~\eqref{eq: flowquet edge dispersive shifts}, as well as with a numerical Floquet calculation for three representative values of the detuning $\delta \equiv \Omega_z/\Omega_R - 1$ and $n=1$. 
Away from resonances, the analytical curves closely follow the numerical results over a broad range of driving frequencies. The principal structures in Fig.~\ref{fig:chi_F} can be understood directly from Eq.~\eqref{eq:chi_f}, whose two contributions possess poles at $\Omega=\left|\Omega_-\right|$ and $\Omega=\Omega_+$. The low- and high-frequency expansions approach the arbitrary-frequency curves in the corresponding limits, while their continuation across the complete frequency interval emphasizes their asymptotic character. In particular, neither limiting expression reproduces the resonant structures or the intermediate-frequency behavior captured by Eq.~\eqref{eq:chi_f}.

For $\delta>0$, the virtual processes associated with the two terms in Eq.~\eqref{eq:chi_f} can, in the regime of $\Omega_-<\Omega <\Omega_+$, generate equal and opposite spin-dependent frequency corrections. Since $\chi_F$ is defined as the absolute value of the corresponding signed quasienergy difference, the sign change of the latter appears as a cusp touching zero whenever $\Omega=\sqrt{\Omega_-\Omega_+}$. No analogous cancellation occurs for $\delta<0$ since $\Omega_+\Omega_-<0$ and the cancellation frequency is not real.

In the high-frequency limit, the Floquet dispersive shift follows the trend $1/\Omega^2$, %thus yielding $\chi_F(\Omega, n=1) \simeq \frac{2\hbar g^2\Omega_Z}{\Omega^2}$. This
asymptotic suppression that is visible after the last resonance of each curve. Physically, a sufficiently rapid modulation averages the transverse coupling over the driving period, reducing the net spin-dependent quasienergy correction.

The resonant case $\delta = 0$ requires additional care in the low-frequency region. Although the compact expression in Eq.~\eqref{eq:chi_f} remains finite for any nonzero $\Omega$, the generator coefficients in Eq.~\eqref{eq: S1 time dependent example low freq} contain denominator terms such as $\Omega_-+\Omega=\Omega$ and $\Omega_- - \Omega=-\Omega$. In this resonant regime, the magnitude of the corresponding generator therefore scales as $g/\Omega$. Consequently, for $\Omega\lesssim g$, the transformation is no longer perturbative, even though the divergent contributions cancel in the final second-order expression for $\chi_F$. With the value of $g=0.1\Omega_R$ employed in Fig.~\ref{fig:chi_F}, this loss of perturbative control occurs precisely in the low-frequency region, where the green numerical points depart from the analytical curve.

%%%%%%%%%%%%%%%%%%%%%%%%%%%%%%%%%%%%%%%%%%%%%%%%
%%%%%%%%%%%%%%%%%%%%%%%%%%%%%%%%%%%%%%%%%%%%%%%%
%%%%%%%%%%%%%%%%%%%%%%%%%%%%%%%%%%%%%%%%%%%%%%%%
%%%%%%%%%%%%%%%%%%%%%%%%%%%%%%%%%%%%%%%%%%%%%%%%
%%%%%%%%%%%%%%%%%%%%%%%%%%%%%%%%%%%%%%%%%%%%%%%%%%%%%%%%%%%%%%%%%%%%%%%%%%%%%%%%%%%%%%%%%%%%%%%%
%%%%%%%%%%%%%%%%%%%%%%%%%%%%%%%%%%%%%%%%%%%%%%%%
\section{Summary} \label{sec: Summary}
In summary this article introduces a novel, systematic approach to constructing the generator of arbitrary perturbative unitary transformations. With this new methodology we offer a general closed-form solution for constructing effective Hamiltonians without the need for heuristic assumptions or the truncation of Hilbert spaces, making it applicable to both finite, infinite-dimensional systems and their combinations. Extending the framework %is also extended 
to accommodate time-dependent systems, we provide a closed-form 
%for which we derived a
solution for time-periodic perturbations (with extension to Sambe space in Appendix~\ref{appendix: sambe space}).

%The chosen examples further demonstrate the reach of this formalism in treating otherwise non-trivially treatable systems using 
To exemplify the advantages of our framework with respect to existing
conventional perturbative constructions, we analyze the dispersive regime of Rabi-like models. 
In  Sec.~\ref{sec: example 1 system},
we study a time independent quantum Rabi model with an anharmonic oscillator.
The presented method allows to fully include the anharmonicity  within the unperturbed Hamiltonian, so that its full occupation-number dependence enters the transformation. This results in a more accurate  effective dispersive interaction, as demonstrated by our spectral analysis and the corresponding dispersive shifts.
%without requiring an additional perturbative expansion. 
In  Sec.~\ref{sec: time dependent example}, we turn the coupling strength of the quantum Rabi model time periodic
and use the close-form solution to derive
 an effective dispersive Hamiltonian valid for  arbitrary driving-frequency. This method is superior to the usual adiabatic or high-frequency treatments, as demonstrated by the accuracy of the computed Floquet dispersive shift.

The presented method to obtain closed-form generators for perturbative unitary transformations  has proven to be a versatile and effective tool for the study of otherwise complicated  static and periodically-driven quantum mechanical systems.  Exploiting the modular nature of the presented solutions, we have implemented these results in a systematic software library~\cite{SymPT}. 
This tool, along with existing libraries~\cite{pymablock}, will serve researchers from different fields by providing a reliable, extensible platform for both analytical derivations and numerical explorations.
%%%%%%%%%%%%%%%%%%%%%%%%%%%%%%%%%%%%%%%%%%%%%%%%
%%%%%%%%%%%%%%%%%%%%%%%%%%%%%%%%%%%%%%%%%%%%%%%%
\section{Acknowledgements} 
We acknowledge useful discussions with Maximilian Rimbach-Russ. GFD and MB acknowledge  funding from the Emmy Noether Programme of the German Research
Foundation (DFG) under grant no. BE 7683/1-1.

LR and MB acknowledge financial support from the University of Augsburg through seed funding project 2023-26.

\appendix
\section{Deriving general operator forms} \label{sec: appendix proof of general operator}
In this section we provide a formal proof for the form of a general hermitian and antihermitian operator provided in Eq.~\eqref{eq: general operator}.
Consider a composite Hilbert space $\mathds{H}$ decomposable as $\mathds{H}_f \otimes \mathds{H}_b$ where $\mathds{H}_f$ is a finite-dimensional Hilbert space of dimensionality $d_f$, while $\mathds{H}_b$ is bosonic subspace with an associate number operator $N = a^\dagger a$. Now consider operators  acting on $\mathds{H}$ as matrices with four “indices” $(\mu,n;\,\nu,m)$.  Any operator $\mathcal{O}$ on $\mathds{H}$ can be written in the “double basis" as
\begin{align}
    \hat{\mathcal O}   =   \sum_{\mu,\nu=0}^{d_f-1}  \sum_{m,n=0}^\infty
O_{\mu,m;\,\nu,n}  \bigl(\ket\mu \bra\nu\otimes\ket m \bra n\bigr).
\end{align}

We now reorganize that double sum by the difference
\begin{align}
    \Delta = n - m\in\mathbb Z\,,
\end{align}
so that $n=m+\Delta$. Let us start by considering the cases $\Delta\geq0$
\begin{align}
\sum_{\mu,\nu}\sum_{\Delta\geq 0} \sum_{m}
O_{\mu,m;\,\nu,m+\Delta}\left(\ket\mu\bra\nu\otimes\ket{m} \bra {m+\Delta}\right).    \label{eq: very general operator in appendix}
\end{align}
At this stage we employ the identity
\begin{align}
    \ket{m}\bra{m+\Delta} & = \sqrt{\frac{m!}{(m+\Delta)!}}\ket{m}\bra{m}\hat a^{\Delta}\\
    %%%%%%
    &=\ket{m}\bra{m}\left[\frac1{\sqrt{(\hat N+1)\cdots(\hat N+\Delta)}}\,\right]\hat a^{\Delta}.
\end{align}
Equivalently, one can absorb the whole $\hat N$ dependent factor into a single operator‐valued function $\hat G_m^{(\Delta)}(\hat N)$. Hence
\begin{align}
    \ket{m}\bra{m+\Delta} =\hat G_m^{(\Delta)}(\hat N)\hat a^{\Delta},    \label{eq: extracting g in appendix}
\end{align}
Plugging back Eq.~\eqref{eq: extracting g in appendix} into Eq.~\eqref{eq: very general operator in appendix} and accounting for its hermitian-conjugate components we get
\begin{widetext}
\begin{align}
\hat{\mathcal{O}}^\pm &=\sum_{\mu,\nu=0}^{d_f-1}\sum_{\Delta\geq0} \underbrace{\left[\sum_{m=0}^\infty O_{\mu,m;\nu,m+\Delta}\,\hat G_m^{(\Delta)}(\hat N)\right]}_{\hat o_{\mu\nu}^{(\hat \Delta)}(\hat N)}\hat a^{\Delta}  \hat \sigma_{\mu\nu}\pm \text{h.c.}\\
%%%%%%%%%%%%%%%%%%%%%%%%
&=\sum_{\mu,\nu=0}^{d_f-1}\sum_{\Delta\geq0} \hat o_{\mu\nu}^{(\Delta)}(\hat N)\hat a^{\Delta}  \hat \sigma_{\mu\nu}\pm \text{h.c.}
\end{align}
\end{widetext}
where, we implicitly note that the $\pm \rm h.c.$ in the above expression is to be applied only to $\Delta>0$ terms. The $\Delta=0$ sector is already complete, because the sum runs over all ordered pairs $(\mu,\nu)$. This concludes the derivation for the form of Eq.~\eqref{eq: general operator}.

%%%%%%%%%%%%%%%%%%%%%%%%
%%%%%%%%%%%%%%%%%%%%%%%%
%%%%%%%%%%%%%%%%%%%%%%%%
\section{Multiple finite and bosonic subspaces} \label{sec: appendix multiple bosonic subspaces}
In this section, we derive a general expression for the $S$ generator for the scenarios in which the system's total Hilbert space can be decomposed as $\mathds{H} = \left(\bigotimes_{j=0}^{M_b-1}\mathds{H}_b^{(j)}\right) \otimes \left(\bigotimes_{k=0}^{M_f-1}\mathds{H}_f^{(k)}\right)$, where $\mathds{H}_b^{(j)}$ are bosonic Hilbert spaces, while $\mathds{H}_f^{(k)}$ are finite Hilbert spaces each with dimensionality $d_f^{(k)}$. Following on the steps of Sec.~\ref{sec: General solution to S generator}, we express the unperturbed Hamiltonian $\mathcal{H}^{(0)}$ for these systems as
\begin{align}
    \hat{H}^{(0)} = \sum_{\vec\mu \in \mathcal{M}} f_{\vec\mu}(\vec{N}) \sigma_{\vec\mu\vec\mu}, \label{eq: general unperturbed hamiltonian appendix}
\end{align}
where $\vec\mu\equiv\left(\mu^{(0)},\mu^{(1)},\dots,\mu^{(M_f-1)}\right)$ and where $\mathcal{M}$ is the set containing all possible permutations of the vector $\vec\mu$

\begin{align}
    \mathcal{M} \equiv \left\{\scalemath{0.81}{\left(\mu^{(0)},\dots,\mu^{(M_f-1)}\right): \mu^{(k)} \in [0, d_f^{(k)} - 1]}\right\}.
\end{align}

Additionally, $\sigma_{\vec\mu\vec\nu} \equiv |\vec\mu\rangle \langle\vec\nu|$ represents the transition operators acting on the basis states $\{|\vec\mu\rangle\}$. The term $ f_{\vec\mu}(\vec{N}) $ is instead a general function of the number operators $\vec{N}=(N_0,N_1,...,N_M)$ of the bosonic subspaces comprising $\mathds{H}$. 
On the other hand, we here postulate that any general hermitian (antihermitian) operator $\hat{\mathcal{O}}^\pm$ acting on $\mathds{H}$ can be reformulated as 
\begin{align}
    \hat{\mathcal{O}}^\pm = \sum_{\vec\mu,\vec\nu \in \mathcal{M}}\sum_{\vec{\Delta}} g_{\vec\mu\vec\nu}^{(\boldsymbol{\Delta})}(\vec{N}) \boldsymbol{a}^{\boldsymbol{\Delta}} \sigma_{\vec\mu\vec\nu} \pm \text{h.c.}, \label{eq: general operator appendix}
\end{align}
where we introduced the shorthand notation
\begin{align}
    \boldsymbol{a}^{\boldsymbol{\Delta}} \equiv \bigotimes_{j=0}^{M_b-1} {a_j^\dagger}^{\alpha_j}a_j^{\beta_j}
\end{align}

where $\Delta_j \equiv (\alpha_j,\beta_j)$, with $\alpha_j,\beta_j\in\mathbb{N}_{\geq 0}$ satisfying $\alpha_j\beta_j=0$, denotes the multi-index specifying the factor $(a_j^\dagger)^{\alpha_j}a_j^{\beta_j}$ acting on the $j$-th bosonic Hilbert space. Products containing both $a_j^\dagger$ and $a_j$ have been normal ordered, with the corresponding number operators absorbed into the coefficient $g_{\vec\mu\vec\nu}^{(\boldsymbol{\Delta})}(\vec{N})$.

From the expression $\left[\hat{H}^{(0)},\hat S^{(j)}\right]=\hat P^{(j)}$, where $\hat P^{(j)}$ is a perturbation of order $j$ represented by an hermitian operator (see Sec.~\ref{sec: framework}), and using the commutation relation $\boldsymbol{a}^{\boldsymbol{\Delta}} f(\vec{N}) = f(\vec{N} - \vec{\alpha} +\vec{\beta}) \boldsymbol{a}^{\boldsymbol{\Delta}}$, we arrive at a general expression for the anti-hermitian generator $S^{(j)}$ 
\begin{align}
    s_{\vec\mu\vec\nu}^{(\boldsymbol{\Delta})}(\vec{N}) = -\frac{1}{\omega_{\vec\mu\vec\nu}^{(\boldsymbol{\Delta})}} p_{\vec\mu\vec\nu}^{(\boldsymbol{\Delta})}(\vec{N}),\label{eq: generator for time independent gamma appendix}
\end{align}
where 
\begin{align}
    \omega_{\vec\mu\vec\nu}^{\boldsymbol{\Delta}} = f_{\vec\nu}(\vec{N}-\vec{\alpha} +\vec{\beta}) - f_{\vec\mu}(\vec{N}) .\label{eq: frequency operator appendix}
\end{align}

%%%%%%%%%%%%%%%%%%%%%%%%%%%%%%%%%%%
%%%%%%%%%%%%%%%%%%%%%%%%%%%%%%%%%%%
%%%%%%%%%%%%%%%%%%%%%%%%%%%%%%%%%%%
\section{Interaction picture method equivalence} \label{sec: appendix proof equivalence interaction}
In this section, we provide a formal derivation of the equivalence between the method provided in~\cite{SW_quantum_information_processing} and Eq. (\ref{eq: generator for time independent gamma}). To ease the computations, here we will focus on systems comprised of a single bosonic and a finite Hilbert subspace; however, note that an analogous derivation can be formulated for more general cases (see Appendix \ref{sec: appendix multiple bosonic subspaces}).

To showcase the equivalence between two methods we start by replacing $\hat{H}^{(0)}$ (see Eq. (\ref{eq: general unperturbed hamiltonian})) and  $\hat P^{(j)}$ (see Eq. (\ref{eq: Pj})) into Eq. (\ref{eq: interaction picture solution})

\begin{widetext}
\begin{align}
    \hat S^{(j)} &= \lim_{t^\prime\to 0}\left[i\int \sum_{\mu,\nu=0}^{d_f-1}e^{i\hat f_\mu(N\hat )t^\prime}\hat \sigma_{\mu\mu}\left(\sum_{m,n=0}^{d_f-1}\sum_{\Delta \ge 0} \hat p_{mn}^{(\Delta)}(\hat N) \hat a^{\Delta} \hat \sigma_{mn} + \text{h.c.}\right)e^{-i\hat f_\nu(\hat N)t^\prime}\hat \sigma_{\nu\nu}dt^\prime\right]\\
    %%%%%%%%%%%%%%%%%%
    &= \lim_{t^\prime\to 0}\left[i\sum_{\mu,\nu=0}^{d_f-1}\sum_{m,n=0}^{d_f-1}\sum_{\Delta \ge 0}\int e^{i\hat f_\mu(\hat N)t^\prime}\hat \sigma_{\mu\mu}\hat p_{mn}^{(\Delta)}(\hat N) \hat a^{\Delta} \hat \sigma_{mn}e^{-i\hat f_\nu(\hat N)t^\prime}\hat \sigma_{\nu\nu}dt^\prime\right] - \text{h.c.}\\
    %%%%%%%%%%%%%%%%%%%%55
    &= \lim_{t^\prime\to 0}\left[i\sum_{\mu,\nu=0}^{d_f-1}\sum_{\Delta \ge 0}\int e^{i\hat f_\mu(\hat N)t^\prime}\hat p_{\mu\nu}^{(\Delta)}(\hat N) \hat a^{\Delta}e^{-i\hat f_\nu(\hat N)t^\prime}\hat \sigma_{\mu\nu}dt^\prime\right] - \text{h.c.}\\
    %%%%%%%%%%%%%%%%%%%%%
    &= \lim_{t^\prime\to 0}\left[i\sum_{\mu,\nu=0}^{d_f-1}\sum_{\Delta \ge 0}\int e^{i\left(\hat f_\mu(\hat N)-\hat f_\nu(\hat N + \Delta)\right)t^\prime}\hat p_{\mu\nu}^{(\Delta)}(\hat N) \hat a^{\Delta}\hat \sigma_{\mu\nu}dt^\prime\right] - \text{h.c.}\\
    %%%%%%%%%%%%%%%%%%%%%55
    &=\lim_{t^\prime\to 0}\left[i\sum_{\mu,\nu=0}^{d_f-1}\sum_{\Delta \ge 0}\frac{e^{i\left(\hat f_\mu(\hat N)-\hat f_\nu(\hat N + \Delta)\right)t^\prime}}{i\left(\hat f_\mu(\hat N)-f_\nu(\hat N + \Delta)\right)} \hat p_{\mu\nu}^{(\Delta)}(\hat N) \hat a^{\Delta}\hat \sigma_{\mu\nu}\right] - \text{h.c.}\\
    %%%%%%%%%%%%%%%%%%%%%5
    &=\sum_{\mu,\nu=0}^{d_f-1}\sum_{\Delta \ge 0}\frac{1}{\hat f_\mu(\hat N)-\hat f_\nu(\hat N + \Delta)} \hat p_{\mu\nu}^{(\Delta)}(\hat N) \hat a^{\Delta}\hat \sigma_{\mu\nu} - \text{h.c.}\\
    %%%%%%%%%%%%%%%%%%%%%%%55
    &=\sum_{\mu,\nu=0}^{d_f-1}\sum_{\Delta \ge 0}-\frac{1}{\hat \omega_{\mu\nu}^{(\Delta)}} \hat p_{\mu\nu}^{(\Delta)}(\hat N) \hat a^{\Delta} \hat \sigma_{\mu\nu} - \text{h.c.}\\
    %%%%%%%%%%%%%%%%%%%5
    &=\sum_{\mu,\nu=0}^{d_f-1}\sum_{\Delta \ge 0}\hat s_{\mu\nu}^{(\Delta)} \hat a^{\Delta}\hat \sigma_{\mu\nu} - \text{h.c.}
\end{align}
\end{widetext}

%%%%%%%%%%%%%%%%%%%%%%%%%%%%%%%%%%%%%%%%%%%%%%%%
%%%%%%%%%%%%%%%%%%%%%%%%%%%%%%%%%%%%%%%%%%%%%%%%
%%%%%%%%%%%%%%%%%%%%%%%%%%%%%%%%%%%%%%%%%%%%%%%%
%%%%%%%%%%%%%%%%%%%%%%%%%%%%%%%%%%%%%%%%%%%%%%%%
%%%%%%%%%%%%%%%%%%%%%%%%%%%%%%%%%%%%%%%%%%%%%%%%
%%%%%%%%%%%%%%%%%%%%%%%%%%%%%%%%%%%%%%%%%%%%%%%%
\section{Static effective Hamiltonian for perturbative anharmonicity}\label{appendix: perturbative anharmonicity}
In Sec.~\ref{sec: example 1 system}, the anharmonic contribution $\hbar\alpha\hat N^2$ was incorporated into the unperturbed Hamiltonian. This choice allowed the occupation-dependent transition frequencies to enter the generator exactly and led to the effective Hamiltoinan in Eq.~\eqref{eq: effective hamiltonian complete}. Here we derive, for comparison, the effective Hamiltonian obtained under the assignlement $\alpha\propto\lambda$. 
The total Hamiltonian can be rewritten as
\begin{align}
    &\hat H(\lambda) \equiv \hat H^{(0)} + \lambda\left(\hat H^{(1)}_{\rm d} + \hat H^{(1)}_{\rm o} \right),\\
    %%%%%%%%%%%%%%%%%%%%%%%%%%%%%%%%5
    &\hat H^{(0)} \equiv \hbar\Omega_R\hat N
    + \frac{\hbar\Omega_Z}{2}\hat\sigma_z,
    \label{eq: appendix harmonic reference}\\
    %%%%%%%%%%%%%%%%%%%%%%%%%%%%%%%%%
    &\hat H^{(1)}_{\rm d} \equiv \hbar \alpha \hat N^2,\label{eq: appendix perturbative anharmonicity}\\
    %%%%%%%%%%%%%%%%%%%%%%%%%%%%%%%%%
    &\hat H^{(1)}_{\rm o} \equiv \hbar g
    \left(
        \hat a+\hat a^\dagger
    \right)
    \hat\sigma_x.
    \label{eq: appendix perturbative interaction}
\end{align}
This new partitioning has the effect of altering the definition of the denominator factor in Eq.~\eqref{eq: omega plus toy}

\begin{align}
    \hbar\hat{\omega}^{(1)}_{\mu\nu} &= \hat{f}_\nu(\hat{N}+1) - \hat{f}_\mu(\hat{N}) \\
&= \hbar\left[\Omega_R +  (-1)^\nu \Omega_Z\right]. \label{eq: omega plus toy appendix}
\end{align}
Inserting Eq.~\eqref{eq: omega plus toy appendix} in Eq.~\eqref{eq: s1_munu}, the first order generator takes the form
\begin{align}
    \hat S^{(1)} = 
    g
    \left[ \hat a\left(
    \frac{1}{\Omega_+} \hat \sigma_{10}
    -\frac{1}{\Omega_-}\hat \sigma_{01}
    \right)
    -
    \rm h.c.
    \right], \label{eq: S1 for QRM}
\end{align}
with $\Omega_\pm \equiv \Omega_Z \pm \Omega_R$, which is the standard generator for the harmonic quantum Rabi model. The effective Hamiltonian then takes the form
\begin{align}
    \mathcal{\hat H} \equiv \hat H^{(0)} + \lambda \hat H^{(1)}_{\rm d} + \frac{\lambda^2}{2} [\hat H^{(1)}_{\rm o}, \hat S^{(1)}].
\end{align}
By neglecting two photon processes terms, as these would be eliminated by a transformation involving a second order generator, we arrive at
\begin{align}
    \hat{\mathcal H} = \hat H^{(0)} + \hat H^{(1)}_{\rm d} + \lambda^2\left[\hat \delta(\hat N) + \hat \xi(\hat N) \hat \sigma_z \right],
\end{align}
where
\begin{align}
    &\hat \delta(\hat N) \equiv \frac{g^2}{2}\left[\frac{1}{\Omega_-} - \frac{1}{\Omega_+} \right],\\
    %%%%%%%%%%%%%%%%%%%%%%%%%%%%%%%%%%%%%%%%%%%%%%%%%%%%%%%%%%%%%%%55
    &\hat \xi(\hat N) \equiv \frac{g^2}{2}\left[\frac{1}{\Omega_-} 
    + \frac{1}{\Omega_+} 
    \right](2\hat N+1).
\end{align}
The dispersive shift reads
\begin{align}
    \chi(n) = 2\hbar g^2 \left|\frac{1}{\Omega_-} 
    + \frac{1}{\Omega_+} \right|n
\end{align}
which, in contrast to Eq.~\eqref{eq: chi(n)}, fails to capture the occupation-number dependence of the transition frequencies induced by the anharmonicity. Since $\hat H^{(1)}_{\rm d}=\hbar\alpha\hat N^2$ does not yet enter the generator at this order, the energy denominators in Eq.~\eqref{eq: omega plus toy appendix} only retain the harmonic content of $\hat H^{(0)}$. Recovering the leading anharmonic correction to $\chi(n)$ within this perturbative partitioning of $\alpha$ would requiring pushing the expansion to third order in $\lambda$, the lowest order at which both $\hat H^{(1)}_{\rm d}$ and $\hat H^{(1)}_{\rm o}$ simultaneously contribute to the generator.

%%%%%%%%%%%%%%%%%%%%%%%%%%%%%%%%%%%%%%%%%%%%%%%%
%%%%%%%%%%%%%%%%%%%%%%%%%%%%%%%%%%%%%%%%%%%%%%%%
%%%%%%%%%%%%%%%%%%%%%%%%%%%%%%%%%%%%%%%%%%%%%%%%
%%%%%%%%%%%%%%%%%%%%%%%%%%%%%%%%%%%%%%%%%%%%%%%%
%%%%%%%%%%%%%%%%%%%%%%%%%%%%%%%%%%%%%%%%%%%%%%%%
%%%%%%%%%%%%%%%%%%%%%%%%%%%%%%%%%%%%%%%%%%%%%%%%

\section{Sambe Space Formalism for Periodic Systems}\label{appendix: sambe space}

While Eq.~\eqref{eq: generator for periodic gamma} provides a complete solution for periodically driven systems in the original time-dependent framework, an elegant alternative emerges through the Sambe space formalism~\cite{Sambe_1973, Sambe_high_frequency}. This approach, particularly powerful for high-frequency drives, transforms the time-dependent problem into an equivalent time-independent one by extending the Hilbert space to include the temporal Fourier modes of the driving field. The resulting framework allows us to apply the static solution of Eq.~\eqref{eq: generator for time independent gamma} directly, bypassing the need to solve the differential equation Eq.~\eqref{eq: generator for time dependent gamma} altogether.

%%%%%%%%%%%%%%%%%%%%%%%%%%%%%%%%%%%%%%%%%%%%%%%%
\paragraph{Extended Hilbert Space and Floquet Shift Operators}
Consider a quantum system subject to a time-periodic Hamiltonian with period $T_{\text{period}}=2\pi/\Omega$. According to Floquet theory, each solution of the Schrödinger equation can be written as
\begin{align}
|\psi_\alpha(t)\rangle
=
e^{-i\varepsilon_\alpha t/\hbar}
|u_\alpha(t)\rangle,
\end{align}
where $\varepsilon_\alpha$ is the quasienergy and the Floquet mode satisfies $|u_\alpha(t+T_{\text{period}})\rangle=|u_\alpha(t)\rangle$. This periodic Floquet mode then admits the Fourier expansion
\begin{align}
|u_\alpha(t)\rangle
=
\sum_{n=-\infty}^{\infty}
|u_{\alpha,n}\rangle e^{-in\Omega t},
\end{align}
with $|u_{\alpha,n}\rangle\in\mathds{H}_\text{phys}$. In the Sambe construction, these Fourier components are promoted to independent coordinates in an extended Hilbert space,
\begin{align}
\mathds{H}_\text{Sambe}
=
\mathds{H}_\text{phys}
\otimes
\mathds{H}_\text{Floquet},
\end{align}
where $\mathds{H}_\text{Floquet}$ is spanned by the orthonormal basis ${|r\rangle}_{r\in\mathbb{Z}}$, satisfying $\langle m|n\rangle=\delta_{mn}$. The integer $r$ labels the temporal harmonic and may be interpreted as the net number of drive quanta exchanged relative to a chosen reference sector.

The mathematical structure of $\mathds{H}_\text{Floquet}$ exhibits a discrete ladder structure similar to a bosonic Fock space, but with fundamental differences: the bilateral infinite-dimensional structure admits unitary shift operators rather than non-unitary creation and annihilation operators. We define a unitary shift operator $\hat{T}$ by its action on the Floquet basis
\begin{align}
    \hat{T} |r\rangle = |r-1\rangle. \label{eq: floquet shift up}
\end{align}
Since $\hat{T}$ is unitary, its adjoint satisfies $\hat{T}^\dagger = \hat{T}^{-1}$ and acts as
\begin{align}
    \hat{T}^\dagger |r\rangle = |r+1\rangle. \label{eq: floquet shift down}
\end{align}
The unitarity condition $\hat{T} \hat{T}^\dagger = \hat{T}^\dagger \hat{T} = 1$ holds automatically from the orthonormality of the basis, and the matrix elements take the simple form $\langle r|\hat{T}^m|r'\rangle = \delta_{r,r'-m}$ for any integer $m \in \mathbb{Z}$ (where $\hat{T}^{-m} = (\hat{T}^\dagger)^m$ for $m > 0$).

We also introduce the Floquet harmonic counting operator $\hat{R}$, which is self-adjoint and diagonal in this basis:
\begin{align}
    \hat{R} |r\rangle = r|r\rangle. \label{eq: floquet number operator}
\end{align}

These operators satisfy commutation relations
\begin{align}
    [\hat{T}, \hat{T}^\dagger] &= 0, \label{eq: floquet commutator TT}\\
    [\hat{R}, \hat{T}] &= -\hat{T}, \label{eq: floquet commutator rT}\\
    [\hat{R}, \hat{T}^\dagger] &= \hat{T}^\dagger. \label{eq: floquet commutator rTdag}
\end{align}
The first relation, $[\hat{T}, \hat{T}^\dagger] = 0$, follows immediately from the unitarity of $\hat{T}$. This starkly contrasts with bosonic operators where $[\hat{a}, \hat{a}^\dagger] = 1$. Nevertheless, the commutators with $\hat{R}$ in Eqs.~\eqref{eq: floquet commutator rT} and \eqref{eq: floquet commutator rTdag} preserve the essential algebraic structure needed for our formalism, encoding how shift operations change the harmonic number.

From these commutation relations, we immediately obtain the fundamental shift identity
\begin{align}
    \hat{T}^m \, \hat{o}(\hat{R}) = \hat{o}(\hat{R} + m) \, \hat{T}^m, \label{eq: floquet shift identity}
\end{align}
for any function $\hat{o}(\hat{R})$ and integer $m \in \mathbb{Z}$. This relation is the Sambe-space analog of the identity $\hat{a}^\Delta \hat{f}(\hat{N}) = \hat{f}(\hat{N}+\Delta) \hat{a}^\Delta$ that proved central to deriving Eq.~\eqref{eq: generator for time independent gamma}, and will serve the same purpose in what follows.

%%%%%%%%%%%%%%%%%%%%%%%%%%%%%%%%%%%%%%%%%%%%%%%%
\paragraph{Sambe Hamiltonian and Perturbative Structure}

For a time-periodic Hamiltonian $\hat{H}(t)$ with Fourier decomposition
\begin{align}
    \hat{H}(t) = \sum_{m=-\infty}^{\infty} \hat{H}_m e^{im\Omega t},
\end{align}
where $\hat{H}_m$ are operators acting on $\mathds{H}_\text{phys}$ and $\hat{H}_{-m} = \hat{H}_m^\dagger$ ensures Hermiticity, the corresponding Sambe Hamiltonian is defined as
\begin{align}
    \hat{H}_\text{Sambe} = \sum_{m=-\infty}^{\infty} \hat{H}_m \otimes \hat{T}^m - \hbar\Omega \, \mathbb{I} \otimes \hat{R}, \label{eq: sambe hamiltonian}
\end{align}
where $\mathbb{I}$ denotes the identity operator on $\mathds{H}_\text{phys}$. The Fourier index $m$ is mapped to the shift operator $\hat{T}^m$ acting on the Floquet subspace, while the term $-\hbar\Omega \hat{R}$ represents the energy cost of occupying different Floquet harmonics: being in the $r$-th sector corresponds to having absorbed or emitted $r$ drive photons, each contributing energy $\hbar\Omega$.

The extended Hamiltonian $\hat{H}_\text{Sambe}$ is manifestly time-independent, and the time-dependent Schrödinger equation $i\hbar \frac{\partial}{\partial t}|\psi(t)\rangle = \hat{H}(t)|\psi(t)\rangle$ becomes equivalent to the eigenvalue problem
\begin{align}
    \hat{H}_\text{Sambe} |\Psi_\text{Sambe}\rangle = \varepsilon |\Psi_\text{Sambe}\rangle,
\end{align}
where the quasi-energy $\varepsilon$ determines the stroboscopic evolution of the system. This mapping transforms our time-dependent perturbation problem into a static one in the extended space, making the machinery of Sec.~\ref{sec: perturbative unitary transformations} directly applicable.

To implement perturbative unitary transformations in Sambe space, we decompose the Hamiltonian according to
\begin{align}
    \hat{H}_\text{Sambe} = \sum_{i=0}\hat{H}_\text{Sambe}^{(i)} + \sum_{j=1} \hat{V}_\text{Sambe}^{(j)},
\end{align}
where the unperturbed part typically takes the form
\begin{align}
    \hat{H}_\text{Sambe}^{(0)} = \hat{H}^{(0)} \otimes \mathbb{I}_\text{Floquet} - \hbar\Omega \, \mathbb{I} \otimes \hat{R}. \label{eq: sambe unperturbed}
\end{align}
Here $\hat{H}^{(0)}$ is the unperturbed Hamiltonian in the physical space, diagonal in the eigenbasis as in Eq.~\eqref{eq: general unperturbed hamiltonian}, and $\mathbb{I}_\text{Floquet}$ denotes the identity in Floquet space. The perturbations $\hat{V}_\text{Sambe}^{(j)}$ encode the time-dependent drive, expanded as
\begin{align}
    V_\text{Sambe}^{(j)} = \sum_{m \neq 0} \hat{V}_m^{(j)} \otimes \hat{T}^m,
\end{align}
where $\hat{V}_m^{(j)}$ represents the $m$-th Fourier component of the $j$-th order time-dependent perturbation in the physical space.

%%%%%%%%%%%%%%%%%%%%%%%%%%%%%%%%%%%%%%%%%%%%%%%%
\paragraph{Universal Solution in Sambe Space}

With the Sambe space structure established, we can now apply the framework of Sec.~\ref{sec: universal solution derivation} to derive a universal solution for the generator in this extended space. Consider our total Hilbert space as $\mathds{H}_\text{Sambe} = \mathds{H}_f \otimes \mathds{H}_b \otimes \mathds{H}_\text{Floquet}$, where $\mathds{H}_f$ and $\mathds{H}_b$ are the finite-dimensional and bosonic subspaces from the physical system. Any Hermitian operator $\hat{\mathcal{O}}^+$ or anti-Hermitian operator $\hat{\mathcal{O}}^-$ in this space admits a decomposition
\begin{align}
    \hat{\mathcal{O}}_\text{Sambe}^\pm = \sum_{\mu\nu} \sum_{\Delta \geq 0} \sum_{m=-\infty}^{\infty} \hat{o}_{\mu\nu}^{(\Delta,m)}(\hat{N}, \hat{R}) \, \hat{a}^{\Delta} \hat{T}^m \hat{\sigma}_{\mu\nu} \pm \text{h.c.},
\end{align}
where the coefficient functions now depend on both number operators $\hat{N}$ (bosonic) and $\hat{R}$ (Floquet).

For the unperturbed Hamiltonian in Eq.~\eqref{eq: sambe unperturbed}, we can write
\begin{align}
    \hat{H}_\text{Sambe}^{(0)} = \sum_{\mu=0}^{d_f-1} [\hat{f}_\mu(\hat{N}) - \hbar\Omega \hat{R}] \, \hat{\sigma}_{\mu\mu},
\end{align}
which is diagonal in the combined basis of finite-dimensional states $\{|\mu\rangle\}$, bosonic Fock states $\{|n\rangle\}$, and Floquet harmonics $\{|r\rangle\}$. Following the same derivation that led to Eq.~\eqref{eq: generator for time independent gamma}, we seek the generator $S_\text{Sambe}^{(j)}$ that satisfies
\begin{align}
    [\hat{H}_\text{Sambe}^{(0)}, \hat{S}_\text{Sambe}^{(j)}] = \hat{P}_\text{Sambe}^{(j)}.
\end{align}

To evaluate the commutator, consider the action on a basis state $|\nu, n, r\rangle$, where $\nu$ labels the finite-dimensional subspace, $n$ is the bosonic occupation number, and $r$ is the Floquet harmonic. A basis operator
\begin{align}
    \hat{\mathcal{O}}_{\mu\nu}^{(\Delta,m)} \equiv \hat{a}^{\Delta} \hat{T}^m \hat{\sigma}_{\mu\nu}
\end{align}
acts by annhilating $\Delta$ bosons, shifting the Floquet index by $m$, and transitioning between finite-dimensional states: $|\nu, n, r\rangle \to |\mu, n-\Delta, r-m\rangle$. The commutator with the unperturbed Hamiltonian therefore produces
\begin{align}
    [\hat{H}_\text{Sambe}^{(0)}, \hat{\mathcal{O}}_{\mu\nu}^{(\Delta,m)}] =- \hbar\hat{\omega}_{\mu\nu}^{(\Delta,m)}(\hat{N}, \hat{R}) \, \hat{\mathcal{O}}_{\mu\nu}^{(\Delta,m)},
\end{align}
where we have defined the energy difference operator
\begin{align}
    \hbar\hat{\omega}_{\mu\nu}^{(\Delta,m)}(\hat{N}, \hat{R}) \equiv \hat{f}_\nu(\hat{N}+\Delta) - \hat{f}_\mu(\hat{N}) - m\hbar\Omega. \label{eq: sambe energy difference}
\end{align}
This operator encodes the energy cost of the transition. More precisely, $\hat{\mathcal O}_{\mu\nu}^{(\Delta,m)}$ maps the initial state $|\nu,N+\Delta,r+m\rangle$ (with energy $f_\nu(N+\Delta)-\hbar\Omega(r+m)$) to the final state $|\mu,N,r\rangle$ (with energy $f_\nu(N)-\hbar\Omega r$).

Expanding both $P_\text{Sambe}^{(j)}$ and $S_\text{Sambe}^{(j)}$ in the operator basis as
\begin{align}
    P_\text{Sambe}^{(j)} &= \sum_{\mu\nu} \sum_{\Delta \geq 0} \sum_{m=-\infty}^{\infty} \hat{p}_{\mu\nu}^{(\Delta,m)}(\hat{N}, \hat{R}) \, \hat{a}^{\Delta} \hat{T}^m \hat{\sigma}_{\mu\nu} + \text{h.c.},\\
    S_\text{Sambe}^{(j)} &= \sum_{\mu\nu} \sum_{\Delta \geq 0} \sum_{m=-\infty}^{\infty} \hat{s}_{\mu\nu}^{(\Delta,m)}(\hat{N}, \hat{R}) \, \hat{a}^{\Delta} \hat{T}^m \hat{\sigma}_{\mu\nu} - \text{h.c.},
\end{align}
and matching coefficients of $\hat{a}^\Delta \hat{T}^m \hat{\sigma}_{\mu\nu}$ on both sides of the defining equation, we obtain
\begin{align}
    -\hbar\hat{\omega}_{\mu\nu}^{(\Delta,m)}(\hat{N}, \hat{R}) \, \hat{s}_{\mu\nu}^{(\Delta,m)}(\hat{N}, \hat{R}) = \hat{p}_{\mu\nu}^{(\Delta,m)}(\hat{N}, \hat{R}).
\end{align}

This leads to our central result: provided that $\hat{\omega}_{\mu\nu}^{(\Delta,m)}(\hat{N},\hat{R})$ is invertible on the relevant state subspace, the generator coefficients are given by
\begin{align}
    \hat{s}_{\mu\nu}^{(\Delta,m)}(\hat{N}, \hat{R}) = -\frac{1}{\hbar\hat{\omega}_{\mu\nu}^{(\Delta,m)}(\hat{N}, \hat{R})}\hat{p}_{\mu\nu}^{(\Delta,m)}(\hat{N}, \hat{R}). \label{eq: sambe generator solution}
\end{align}

This expression constitutes the direct generalization of Eq.~\eqref{eq: generator for time independent gamma} to the Sambe space. This derivation highlights a crucial caveat of the perturbative treatment of driven systems: the inverse of the energy difference operator $\hat{\omega}_{\mu\nu}^{(\Delta,m)}(\hat{N},\hat{R})$, only exists when it has non-zero eigenvalues. When the resonance condition
\begin{align}
    \hat{f}_\nu(\hat N+\Delta) - \hat{f}_\mu(\hat N) - m\hbar\Omega = 0 \label{eq: resonance condition}
\end{align}
is satisfied for some state $|n\rangle$ and transition indices $(\mu, \nu, \Delta, m)$, the formal inversion in Eq.~\eqref{eq: sambe generator solution} fails. Physically, this corresponds to the driving frequency $\Omega$ (or one of its harmonics $m\Omega$) matching a transition frequency in the system, enabling resonant excitation that cannot be treated perturbatively through unitary transformation approach.

\section{Approximate frequencies expansions} \label{appendix: edge frequency example}

The analytical expression obtained in Sec.~\ref{sec: time dependence} for the driven Rabi model remains valid for abitrary driving frequencies provided that the perturbative denominators remain sufficiently far from the systems resonances. It is nevertherless useful to examine the limiting forms of the effective Hamiltonian, both to connect the general result with an adiabatic and a high frequency treatement and to clarify the asymptotic curves displayed in Fig.~\ref{fig:chi_F}.

\subsection{Low-frequency limit of the driven system} \label{appendix: low frequency example}
We first consider the low frequency regime, defined by $|\Omega|\ll|\Omega_\pm|$. Within this limit, the perturbative transformation varies slowly enough that the $\frac{\partial \hat S^{(1)}(t)}{\partial t}$ is of higher perturbative order than $\hat S^{(1)}(t)$ itself. As discussed in Sec.~\ref{sec: time dependent transformations}, it follows that the defining equation for the generator $\hat S^{(1)}(t)$ is no longer Eq.~\eqref{eq: time dependent defining equation}, but instead by its time independent variant \eqref{eq: Equation to solve}. By following the considerations carried out in Sec.~\ref{sec: time dependent example}, namely
\begin{align}
    \hat p_{\mu\nu}^{(1)} =- \hbar g\cos(\Omega t),
\end{align}
the generator coefficients are obtained via Eq.~\eqref{eq: generator for time independent gamma} 

\begin{align}
    \hat s^{(1)}_{01}(t)
    =
    -\frac{g}{\Omega_-}\cos(\Omega t),
    \quad
    \hat s^{(1)}_{10}(t)
    =
    \frac{g}{\Omega_+}\cos(\Omega t).
    \label{eq: low frequency generator coefficients}
\end{align}

The corresponding first-order generator therefore reduces to 
\begin{align}
    \hat S^{(1)}(t) = 
    g\cos(\Omega t)
    \left[ \hat a\left(
    \frac{1}{\Omega_+} \hat \sigma_{10}
    -\frac{1}{\Omega_-}\hat \sigma_{01}
    \right)
    -
    \rm h.c.
    \right].
    \label{eq: low frequency generator}
\end{align}
Equation~\eqref{eq: low frequency generator} has the same operator structure as the generator of the static example studied in Sec.~\ref{sec: example 1 system} for $\alpha=0$, with the coupling amplitude replaced by its instantaneous value $g\cos(\Omega t)$. This is the expected behaviour in the low-frequency regime, where the perturbative transformation follows the modulation quasi-statically, as discussed in Sec.~\ref{sec: time dependence}. 

The effective Hamiltonian can be obtained by substituting Eq.~\eqref{eq: low frequency generator} into the second-order commutator in Eq.~\eqref{eq: TD effective hamiltonian complete}. Equivalently, and more directly, one may expand the frequency-dependent coefficient appearing in that equation, arriving at
\begin{align}
    \hat{\mathcal H}(t)
    =
    \hat H^{(0)}
    +
    \lambda^2
    \frac{\hbar g^2\Omega_Z}{\Omega_Z^2-\Omega_R^2}
    \cos^2(\Omega t)
    \left(2\hat N+1\right)
    \hat\sigma_z.
    \label{eq: low frequency effective Hamiltonian}
\end{align}
The low-frequency approximation has therefore reduced the driven problem to the usual dispersive Hamiltonian of the static Rabi model, modulated by the instantaneous squared coupling amplitude. The time dependence remains relevant for the micromotion, but the Hamiltonian in Eq.~\eqref{eq: low frequency effective Hamiltonian} continues to commute with itself at different times. Its one-period evolution may thus be evaluated by direct integration, as was done for the complete result in Eq.~\eqref{eq: TD one period evolution}. Setting $\lambda=1$ one finds the low-frequency Floquet Hamiltonian 
\begin{align}
    \hat{\mathcal H}_F^{\rm low}
    =
    \hat H^{(0)}
    +
    \hbar g^2
    \frac{\Omega_Z}{\Omega_Z^2-\Omega_R^2}
    \left(
        \hat N+\frac{1}{2}
    \right)
    \hat\sigma_z.
    \label{eq: low frequency Floquet Hamiltonian}
\end{align}

Comparison with the general form introduced in Eq.~\eqref{eq: TD one period evolution} identifies the dispersive shift as
\begin{align}
    \chi_F^{\rm low} = 2\hbar g^2\left|
    \frac{\Omega_Z}{\Omega_Z^2- \Omega_R^2} 
    \right|n 
    \label{eq: low frequency chi F}.
\end{align}
At the retained order, Eq.~\eqref{eq: low frequency chi F} is independent of the driving frequency. It consequently appears as the horizontal low-$\Omega$ asymptote in Fig.~\ref{fig:chi_F} .

%%%%%%%%%%%%%%%%%%%%%%%%%%%%%%%%%%%%%%%%%%%%%%%%
%%%%%%%%%%%%%%%%%%%%%%%%%%%%%%%%%%%%%%%%%%%%%%%%
\subsection{High-frequency limit of the driven system}
\label{sec: appendix high frequency driven system}
We next consider the complementary regime in which the modulation is fast compared with both transition frequencies of the unperturbed system, such that $|\Omega|\gg |\Omega_\pm|$. This limit connects the general transformation developed in Sec.~\ref{sec: time dependence} with the standard structure of a high-frequency Floquet-Magnus expansion~\cite{SW_TD_FloquetSW}. 

To derive such high frequency expansion, let us partition the perturbation term in Eq.~\eqref{eq: driven rabi Vt} in its harmonic decomposition
\begin{align}
    &\hat H^{(1)}(t) = \hat H_{+1}^{(1)}e^{i\Omega t} + \hat H_{-1}^{(1)}e^{-i\Omega t} \\
    %%%%%%%%%%%%%%%%%%%%%%%%%%%%%%%%%%%%%%%%%%%%5
    &\hat H_{+1}^{(1)} = \hat H_{-1}^{(1)} = \frac{\hbar g}{2} \left(\hat a^\dagger + \hat a \right)\hat \sigma_x.
\end{align}
Using the above expressions, the Hamiltonian up $\frac{1}{\Omega^2}$ (note that lower order terms vanish) takes the form
\begin{align}
    \hat{\mathcal{H}}_F^{\rm{high}} &= \hat H^{(0)} + \frac{\left[\hat H^{(1)}_{+1}, \left[\hat H^{(0)}, \hat H^{(1)}_{+1},\right]\right]}{(\hbar\Omega)^2}\\
    %%%%%%%%%%%%%%%%%%%%%%%%%%%%%%%%%%%%
    &=\hat H^{(0)} - \hbar g^2\frac{\Omega_Z}{\Omega^2}\left(\hat N + \frac{1}{2}\right)\hat \sigma_z \label{eq: high frequency Floquet Hamiltonian}
\end{align}

Comparison with Eq.~\eqref{eq: TD one period evolution} then gives
\begin{align}
    \chi^{\rm{high}}_F(\Omega, n)
    =
    2\hbar g^2
    \left|
        \frac{\Omega_Z}{\Omega^2}
    \right|n.
    \label{eq: high frequency chi F}
\end{align}
Equation~\eqref{eq: high frequency chi F} shows that the driven dispersive shift vanishes quadratically with increasing driving frequency. This behavior produces the $1/\Omega^2$ high-frequency asymptote displayed in Fig.~\ref{fig:chi_F}.

\bibliography{references}

\end{document}